\documentclass{aa}  

\usepackage{graphicx}
\usepackage{txfonts}

\usepackage{lipsum}
\usepackage{xcolor}
\usepackage{hyperref}
\usepackage[normalem]{ulem}
\usepackage{subcaption}
\usepackage{booktabs}

\newcommand{\ye}{\ensuremath{Y_e}}
\newcommand{\yemin}{\ensuremath{Y_e^\mathrm{min}}}
\newcommand{\msun}{M_\odot}

\newcommand{\mch}{M_\mathrm{Ch}}

\newcommand{\rhoini}{\ensuremath{\rho_c^\mathrm{ini}}}
\newcommand{\modt}[2]{\ensuremath{\mathrm{rho}#1\_\mathrm{r}#2}}
\newcommand{\mods}[2]{\ensuremath{\mathrm{rho}#1\_\mathrm{r}#2\_\mathrm{sfs}}}

\defcitealias{timmes1992a}{TW92}
\defcitealias{schwab2020a}{S20}

\usepackage{hyperref}
\hypersetup{
  colorlinks,
  citecolor=blue,
  linkcolor=blue,
  urlcolor=blue}

\begin{document} 
   \title{Drawing the line between explosion and collapse in electron-capture supernovae}
   \subtitle{I. Impact of conductive flame speeds and ignition conditions on the explosion mechanism}
   \titlerunning{Drawing the line between explosion and collapse in electron-capture supernovae}
    \authorrunning{Holas et al.}

   \author{
        Alexander Holas\inst{1,*}
        \and
        Samuel W. Jones \inst{2,*}
        \and
        Friedrich K. Röpke\inst{1,3}
        \and
        Rüdiger Pakmor\inst{4}
        \and
        Christina Fakiola\inst{1,3,*}
        \and
        Giovanni Leidi\inst{1}
        \and
        Raphael Hirschi\inst{5,6,*}
        \and
        Ken J. Shen\inst{7}
    }

   \institute{
        Heidelberger Institut für Theoretische Studien,
        Schloss-Wolfsbrunnenweg 35, 69118 Heidelberg, Germany\\
        \email{alexander.holas@mailbox.org}
        \and
        Theoretical Division, Los Alamos National Laboratory, Los Alamos, NM 87545, US
        \and
        Zentrum f\"ur Astronomie der Universit\"at Heidelberg,
        Institut f\"ur Theoretische Astrophysik, 
        Philosophenweg 12,
        69120 Heidelberg, Germany
        \and
        Max-Planck-Institut f\"ur Astrophysik,
        Karl-Schwarzschild-Str. 1, 85748 Garching,
        Germany
        \and
        Astrophysics Research Centre, Lennard-Jones Laboratories, Keele University, Keele ST5 5BG, UK
        \and
        Kavli IPMU (WPI), The University of Tokyo, 5-1-5 Kashiwanoha, Kashiwa 277-8583, Japan
        \and
        Department of Astronomy and Theoretical Astrophysics Center, University of California, Berkeley, CA 94720, USA\\
        $^*$The NuGrid collaboration
    }

   \date{Received September 15, 1996; accepted March 16, 1997}    

    \abstract{
    Electron-capture supernovae (ECSNe) are commonly thought to result in a collapse to a neutron star. Recent work has shown that, under certain conditions, a thermonuclear explosion is also a possible outcome.
    The division between the two regimes, however, has not yet been mapped out.
    }{
    In this study, we investigate the conditions under which the transition from thermonuclear explosion to collapse occurs, and what physical mechanisms drive each outcome.
    }{
    We conducted a parameter study of $56$ 3D hydrodynamic simulations of ECSN in ONe white dwarfs using a level set based flame model implemented in the \textsc{Leafs} code.
    We varied both the ignition location and the central density at ignition to determine the conditions of the transition regime.
    Additionally, we explored two different laminar flame parameterizations and how they impact the simulation outcome.
    }{
    From our parameter study, we find a transition density in the range of $\log\rhoini =10.0$ and $10.15\,\mathrm{g}\,\mathrm{cm}^{-3}$, depending on the ignition location and utilized laminar flame speed parameterization.
    Importantly, we find that for sufficiently high central densities, the burned ashes can sink into the core and trap large amounts of neutron-rich material in the bound remnant.
    In the transition regime between explosion and collapse, we find that the laminar flame speed plays a critical role by suppressing the formation of instabilities and thereby reducing the nuclear energy generation needed to overcome the collapse.    
    }{
    We find that a thermonuclear explosion is possible for a wide range of parameters, whereby a more off-center ignition allows for higher central densities to still result in an explosion.
    Both the conditions at ignition and the flame physics are critical in determining the outcome.
    Detailed 3D hydrodynamic simulations of the preceding stellar evolution and the ignition process of the thermonuclear flame are necessary to accurately predict the outcome of ECSNe.
    } 

   \keywords{white dwarfs --
                supernovae: general --
                hydrodynamics --
                Stars: AGB and post-AGB --
                Stars: neutron
               }

   \maketitle
   \nolinenumbers

\section{Introduction}

After decades of research, the explosive death of most stars has been well explored.
While the most massive stars ($\gtrsim12\,\msun$) end their lives in an iron core-collapse supernova (FeCCSN), collapsing under their own weight and resulting in the formation of either a neutron star (NS) or a black hole (BH), stars at the lighter end of the mass spectrum ($\lesssim 8\,\msun$) will form a white dwarf (WD) star.
WDs are typically stable objects when they evolve in an isolated environment, but when interacting with a companion star, the interaction can trigger a thermonuclear explosion commonly known as Type~Ia supernova (SN~Ia; \citealt{whelan1973a,iben1984a}). 
Although there is still an active discussion surrounding the two explosion channels (i.e., the thermonuclear and iron core-collapse channels; see \citealt{liu2023a} and \citealt{janka2025a} for recent reviews), the upper and lower end of the stellar mass range has overall been well explored, and there is at least a superficial consensus on how these stars end their lives.
In contrast, the fate of intermediate mass stars ($8-12\,\msun$) remains equivocal and has so far been sparsely explored.
A common idea is that these stars end their lives as so-called electron-capture supernovae (ECSNe; \citealt{miyaji1987a,nomoto1984a,nomoto1987a}), where, initiated by electron capture reactions, the star loses the pressure support provided by degenerate electrons and collapses under its own weight, resulting in an SN explosion and leaving behind an NS.
Previous research \citep{isern1991a,canal1992a,jones2016a,jones2019b,jones2019a,kirsebom2019a,leung2020a} has shown that, contrary to the previous notion that ECSNe end in a collapse, a thermonuclear explosion can in theory also be achieved.
Although these thermonuclear ECSNe (tECSNe) can help explain some otherwise difficult to produce species in the solar inventory of isotopes --- tECSNe show a strong production of ${^{48}}\mathrm{Ca}$, ${^{50}}\mathrm{Ti}$, and ${^{52}}\mathrm{Cr}$ as well as several Zn-Zr isotopes \citep{jones2019a} --- without introducing substantial tensions elsewhere, it is not clear if tECSNe exist in nature or if they are purely a theoretical phenomenon.
While synthetic observables of collapsing ECSNe (cECSNe) have been matched to existing observations \citep[e.g.,][]{kozyreva2021a}, they differ little from low-mass CCSNe, and the observational properties of tECSNe are so far unknown.
Moreover, it is not yet clear what determines either outcome and how the transition from collapse to thermonuclear explosion impacts observational properties such as the ejecta composition.
Answering the latter question is the focus of the current work.

In general, ECSNe occur when the ONe core of an intermediate mass star --- commonly super asymptotic giant branch (sAGB) stars (see \citet{doherty2017a} for a review) --- reaches the Chandrasekhar mass limit ($\mch$), usually through repeated unstable He shell burning cycles \citep{ritossa1999a,jones2013a}.
This mass limit can also be reached in an ONe WD that stably accretes mass from a binary companion \citep{schwab2015a}, also known as accretion-induced collapse (AIC; see \citealt{nomoto1991a}), or in the case of a close binary system through stable He shell burning after the envelope of the star has been stripped \citep{podsiadlowski2004a,tauris2015a}.
Since the precise progenitor scenario is somewhat ambiguous, we are somewhat agnostic of the precise channel by which $\mch$ is reached.

Regardless of the mass accretion mechanism, once the ONe core reaches a critical central density, the electron capture reaction ${^{20}}\mathrm{Ne}\rightarrow {^{20}}\mathrm{F}\rightarrow {^{20}}\mathrm{O}$ will set in.\footnote{We note that depending on the precise stellar evolution, other electron capture reactions may occur before this point that can trigger thermonuclear burning; for example, electron captures on ${^{24}}\mathrm{Mg}$ can ignite residual C (see, e.g., \citealt{jones2013a,schwab2019b,antoniadis2020a,chanlaridis2022a}).}
This reaction has a twofold effect.
On the one hand, it causes the pressure support of the core to drop and therefore the core to contract, further increasing its density and the rate at which electron captures occur.
This initiates a feedback loop that, if nothing else can counteract it, will cause the core to collapse.
On the other hand, this electron capture chain releases energy through $\gamma$-decay of ${^{20}}\mathrm{O}$ \citep{martinez-pinedo2014a} and thereby can ignite runaway thermonuclear burning of the ONe plasma in the form of ${^{16}}\mathrm{O} + {^{16}}\mathrm{O}$ burning \citep{isern1991a,canal1992a,gutierrez1996a,schwab2015a,schwab2017a,zha2019a}.
Although the resultant increase in temperature will increase the electron capture rate on ${^{20}}\mathrm{Ne}$ (as well as other isotopes found in the burning products) and thereby accelerate the collapse, thermonuclear burning provides a mechanism to counteract the onset of collapse. 
In this pivotal moment, the outcome of an ECSN depends on which process can outweigh the other.

Typical cECSNe simulations \citep[e.g.,][]{janka2008a,wanajo2011a,zha2022a} predict that eventually the gravitational collapse will overcome the thermonuclear burning, and the resulting cECSNe will lead to the formation of an NS.
\citet{jones2016a} found that turbulent burning plays an important role in the evolution of an ECSNe and can potentially be a critical factor in halting the gravitational collapse and subsequently overturning it, leading to a thermonuclear explosion.

From previous studies, both outcomes have been established.
However, it is not clear where the transition from one regime to the other occurs.
Specifically, it remains uncertain how the central density that the ONe core reaches at the onset of thermonuclear runaway, along with the location of ignition (or more precisely the impact of convection prior to the ignition; see, e.g., \citealt{gutierrez1996a}), influences the final outcome.
Although there exist 1D studies investigating the fundamentals of flame propagation in this regime (e.g., \citealt{timmes1992a,schwab2020a}), as well as 2D studies of various initial conditions \citep[e.g.,][]{leung2020a,zha2022a}, this issue has not yet been studied in 3D hydrodynamic simulations.
This is important because the fate of the star critically depends on the competition between energy release in the thermonuclear burning and collapse, where the former effect is determined by intrinsically 3D fluid dynamical instabilities that interact with the flame.

We conduct a parameter study, investigating how the central density at ignition and the ignition location impact the outcome of ECSNe.
With this study, we aim to establish what conditions support a tECSN outcome and when a cECSN occurs and specifically what physical mechanisms determine either outcome.
We also investigated the impact of the laminar flame speed parameterization utilized by our simulations on the evolution of the modeled ECSNe.
Furthermore, with our study we aim to provide a large model set that can be used in future studies to establish observational properties of tECSNe in order to determine whether these objects exist in nature or if they remain in the realm of theory.

Our work is organized as follows.
First, we present the numerical tools used in our parameter study as well as the included physics in Section~\ref{sec:methods}.
Second, we provide an overview over the different ECSN modes observed in our models in Section~\ref{sec:ecsn_modes}.
Third, Section~\ref{sec:flamespeeds} provides a detailed investigation into the impact of the laminar flame speed as well as what role turbulence plays in overcoming the collapse.
Finally, Section~\ref{sec:paramstudy} presents the results of our parameter study itself and a discussion about under which conditions our simulations yield a tECSN or cECSN outcome.
We end with some concluding remarks and an outlook on future work in Section~\ref{sec:conclusion}.

\section{Methods}~\label{sec:methods}

\subsection{Explosion simulations}\label{sec:leafs}
Our ECSN simulations were performed using the \textsc{Leafs} code.
This code has been used in a variety of SN~Ia-like applications (see, e.g., \citealt{ropke2005a,ropke2007a,seitenzahl2013a,fink2014a,fink2018a,ohlmann2014a,marquardt2015a,lach2022b,lach2022a}), and, in particular, other ECSN studies \citep{jones2016a,jones2019a,kirsebom2019a}.
The \textsc{Leafs} code has been adapted from the \textsc{Prometheus} code \citep{fryxell1989a} by \cite{reinecke1999a} for its use in SN~Ia studies.
It uses a piecewise parabolic scheme \citep{colella1984a} to solve the reactive Euler equations.
The computation takes place on a 3D Cartesian grid that is able to follow both the evolution of the flame front and the expansion of the star by nesting a uniform Cartesian grid inside a nonuniform Cartesian grid \citep{ropke2006a}.
Flame fronts were treated using a level-set technique \citep{osher1988a} as described by \cite{reinecke1999b}.
Our fiducial set of simulations uses the laminar flame speed prescription of \citet[hereafter TW92]{timmes1992a}; for select models, we also compare simulations using the more recent flame speed parameterization of \citet[hereafter S20]{schwab2020a}.
Both \citetalias{timmes1992a} and \citetalias{schwab2020a} obtain their parameterization from a linear fit to simulations of conductive flames under various conditions; the latter, however, includes more modern microphysics and a larger nuclear reaction network.
In particular, the formula of \citetalias{schwab2020a} considers a varying value of $Y_e$ for a fixed fuel composition, whereas the formulation of \citetalias{timmes1992a} assumes $Y_e=0.5$ and varies the fuel composition.

A subgrid model for turbulence \citep{schmidt2006a,schmidt2006b} was used to account for the contribution of unresolved turbulent velocity fluctuation to the effective flame speed.
In contrast to the work of \cite{jones2016a,jones2019a}, we did not use a spherical Newtonian gravitational potential, but rather a fast Fourier transform-based approach, solving the full Poisson equation for Newtonian gravity.
During the initial phase of the explosion, we did not expect significant deviations from a spherical potential.
However, because pure deflagrations can leave behind bound remnants with substantial kick velocities \citep[e.g.,][]{lach2022b}, thereby deviating from spherical symmetry, we opted for the more flexible approach of solving the full Poisson equation.

The code uses an equation of state (EOS) that comprises a mixture of electrons and positrons modeled as an arbitrarily relativistic Fermi gas, ions modeled as a perfect gas, and thermal radiation, following the implementation described in \cite{timmes2000a}.\footnote{The ECSN simulations of \cite{jones2016a,jones2019a} used the Timmes EOS \citep{timmes1999a} and Coulomb corrections as formulated by \citet{potekhin2000a} instead.}
All of our simulations include Coulomb corrections, following the formulation of \cite{yakovlev1989a}.
Here we utilized a Butterworth filter \citep{butterworth1930a} to allow for a smoother transition between the corrected and non-corrected regimes to avoid numerical issues arising from the sharp transition found in the original implementation.\footnote{\url{https://cococubed.com/codes/eos/helmholtz.tbz}}
Nuclear burning inside the flame front is included through tabulated abundances of the burning products (see \citealt{fink2010a,ohlmann2014a,jones2016a}).
This approximate nuclear network estimates the energy release from nuclear reactions by following six pseudo-species: ${^4}\mathrm{He}$, ${^{12}}\mathrm{C}$, ${^{16}}\mathrm{O}$, ${^{20}}\mathrm{Ne}$, ${^{28}}\mathrm{Si}$, and ${^{56}}\mathrm{Ni}$.
This also includes thermal neutrino losses using the formulae of \citet[see \citealt{seitenzahl2015a}]{itoh1996a}.

We followed the (de)leptonization of the NSE ashes as described in \citet[section 2.3]{jones2016a}.
Here, electron captures and in general reactions such as $\beta$-decays and $e^+$-captures are accounted for; their impact on $Y_e$ is calculated from tabulated $\dot{Y}_e(\rho,T,Y_e)$ values.
This table was computed by \citet{jones2016a} following \citet{seitenzahl2009a} and \citet{pakmor2012a}.
The relevant rates were taken from \citet{langanke2000a}, \citet{oda1994a}, \citet{fuller1985a}, and \citet{nabi2004a}, with approximations from \citet{arcones2010a} and \citet{sullivan2015a}.
This formalism also accounts for the mean energy losses from neutrinos produced in these reactions.
We note that we assumed that if $Y_e \lesssim 0.25$ is reached in the plasma, the density must be high enough for neutrino interactions to substantially alter $Y_e$. 
We do not account for neutrino interactions in our work and assumed that our simulations become invalid as $Y_e\rightarrow0.25$.
Therefore, we limited our $\dot{Y}_e$ table to $Y_e \geq 0.25$.

Lastly, to increase the robustness of the numerical scheme, we extended the original method of \citet{colella1984a} by switching to a more dissipative Rusanov flux function \citep{rusanov1962a, badwaik2020a} in regions where high-Mach flows develop, such as during the interaction of the SN ejecta with the background pseudo-vacuum.
This pseudo-vacuum was set to a fixed density ($1\times10^{-5}\,\mathrm{g}\,\mathrm{cm}^{-3}$) and energy ($1.0\,\mathrm{erg}$) and is therefore not in hydrostatic equilibrium.

\subsection{Initial conditions}
Our initial ONe WD models\footnote{Our simulations neglect the impact of a potential stellar envelope. Given the spatial extent of such an envelope (see \citealt{jones2013a}), it will have little effect on the explosion mechanism, and therefore we only focus on the explosion of the bare core (or a WD).} closely follow the initial conditions of \cite{jones2016a,jones2019a}:
We considered a homogeneous composition of 65\% ${^{16}}\mathrm{O}$ and 35\% ${^{20}}\mathrm{Ne}$ by mass, and parameterize the hydrostatic equilibrium configuration of the resulting $\mch$ WD by its central density \rhoini.
Furthermore, we imposed that the initial temperature stratification is flat and the temperature is $5\times10^5\,\mathrm{K}$.
Due to the fully degenerate nature of WD matter, the temperature does not significantly impact the subsequent evolution of the system.
Importantly, temperature dependent electron capture reactions are only simulated inside the burning region, where the temperature is determined by the nuclear energy release; outside this region, electron capture reactions are too slow to play a role on the timescales simulated here.
To account for the preceding stellar evolution, we used a constant $Y_e = 0.493$ for the entire WD.
This value was calculated by \cite{jones2013a} using a mass-weighted average of a $8.75\,\msun$ progenitor computed with the stellar evolution code \textsc{Mesa} \citep{paxton2011a,paxton2013a,paxton2015a}.
For more details and discussion, see \citet{jones2013a,jones2016a}.

In our parameter study, we used several central densities ranging from $9.95 < \log_{10}\rhoini < 10.4$.
This range represents the uncertainty in the ignition density corresponding to the ability (or not) of convection or semi-convection to transport heat away from the hot spot efficiently enough to delay the ignition and allow the star to contract further before the explosion \citep[e.g.,][]{gutierrez1996a}.
This density space is narrowed down for each ignition geometry (see Section~\ref{sec:ign_geom}) in a binary search pattern to narrow the density at which the collapse transitions to a thermonuclear explosion; due to the computational cost, a uniform density grid would be infeasible.
For a complete list of the central densities of the initial models, see Table~\ref{tab:model_list}.\footnote{We note that the simulations were not run in sequence, and therefore some parameter combinations could be skipped, hence, the asymmetric parameter spacing.}

\subsection{Ignition geometries}\label{sec:ign_geom}
The ignition process occurs on scales not resolved by our numerical grid, and therefore we had to assume an ignition geometry once the initial flame had grown to a size resolved by our simulation.
This gave us some freedom in choosing the initial shape of the flame.
The location of the flame can in principle be determined by either 1D stellar evolution simulations or, if the electron captures trigger a convective instability, 3D reactive hydrodynamics simulations.
At the time of writing there exist no self-consistent 3D stellar evolution models of such ONe core ignitions that sufficiently constrain the ignition location.
Therefore, guided by 1D stellar evolution simulations with state-of-the-art nuclear physics input \citep{kirsebom2019a}, we examined the dependence of the transition density on the ignition location by adopting five different ignition locations: $2.5$\,km, $10$\,km, $20$\,km, $35$\,km, $50$\,km, and $73$\,km offset from the center of the WD.
In contrast to previous simulations of thermonuclear explosions in CO WDs \citep[e.g.,][]{seitenzahl2013a} and the tECSN simulations of \citet{jones2016a}, we ignited in a single location.
Here, we started our parameter grid at $2.5$\,km instead of $0$\,km, as the latter would lead to artificially symmetric explosions; $2.5$\,km, however, means that the center of the WD is still covered by our ignition bubble (see below), but the symmetry is broken by introducing a preferential direction.
The upper limit of $73$\,km was motivated by the work of \citet{kirsebom2019a} and \citet{zha2019a} who suggest that a single hotspot can form up to around this radius.
We did not investigate multi-spot ignitions as they are deemed unlikely \citep{kuhlen2006a,zingale2009a,nonaka2012a}.
For multi-spot ignitions of tECSNe, see \citet{jones2016a}.

As an ignition spot, we chose a composite of nine individual overlapping spheres, similar to the ignition spot utilized by \cite{lach2022b}, effectively adding a initial perturbations on scales resolved by the grid that would naturally develop from Rayleigh-Taylor (RT) instabilities.
Here, each individual sphere has a radius of $5$\,km, and the entire structure has a width of $18.5$\,km.
These values were chosen such that each individual sphere is resolved by at least a few cells, whereas the nine spheres are provided to add some initial perturbations on the resolved scales compared to just one single, larger sphere.\footnote{We note that compared to \cite{lach2022b}, our ignition spot has an additional ninth sphere in the center.}
We found no indication that this choice has a noticeable impact on the simulation results.
In particular, the outcome of our simulations seems robust against the precise shape and size of the ignition bubble.

In each simulation, this composite ignition bubble was placed at the ignition locations listed above.
The simulation grid was tuned so that it retains a high resolution in the central region, leading to a resolution of $700$\,m in the uniform part of the grid (up to a distance of $135$\,km along each axis); this allowed us to at least roughly resolve the curvature of the ignition spheres.
To ensure comparability between simulations, the same initial grid was used throughout the parameter space; for example, for the more central ignitions we could narrow down the central uniform region to achieve higher central resolution, but for the most off-center cases the ignition spot would end up outside this region.
During the first timestep, the material inside the ignition bubble is then burned to ashes while maintaining pressure equilibrium, initiating the explosion without introducing spurious shocks.
As a consequence, the density inside the ignition spot is artificially reduced.
However, the thereby induced buoyant motion does not interfere with the flame front and is, in general negligible compared to the buoyancy naturally developing over time.

We labeled each WD explosion model as rho<central density>\_r<ignition radius>, that is, the model with $\log\rhoini = 9.95\,(\mathrm{g}\,\mathrm{cm}^{-3})$ ignited at $73$\,km would be named \modt{9.95}{73}.
In simulations where we used the updated flame speed prescription of \citetalias{schwab2020a}, we appended the suffix \_sfs; otherwise the flame speeds of \citetalias{timmes1992a} are utilized.
A complete overview of all $56$ simulated models can be found in Table~\ref{tab:model_list}.

\section{ECSNe modes}\label{sec:ecsn_modes}
\begin{figure*}
    \centering
    \includegraphics[width=\textwidth,keepaspectratio]{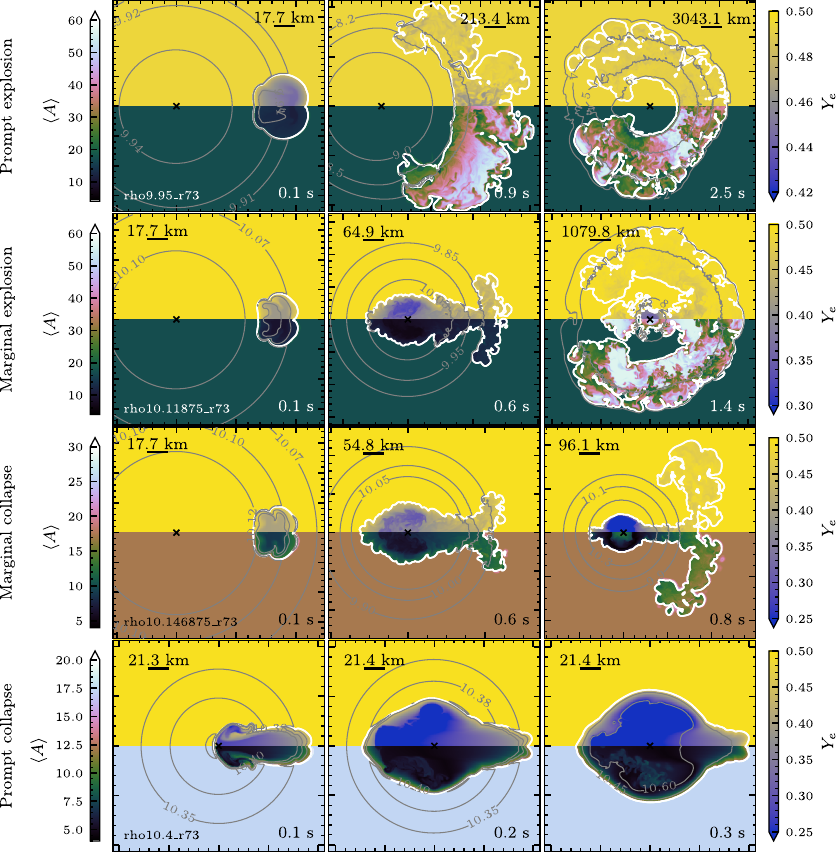}
    \caption{Illustration of the different ECSN modes observed in our parameter study. Each row shows a separate simulation, with each column representing various characteristic times. In each plot, the top half indicates $Y_e$ and the bottom half $\langle A\rangle$ (note the varying limits in the color bars). Here, we also show contours of uniform $\log \rho\,(\mathrm{g}\,\mathrm{cm}^{-3})$. We note that in the last timestep of \modt{10.4}{73}, we no longer fully trust the results of the simulation (e.g., the reappearance of heavier elements in the ashes is likely only an artifact) and only include this timestep as a visual indication of how the ashes get compressed into a sphere.}
    \label{fig:expl_modes}
\end{figure*}

Before investigating the specific conditions under which the thermonuclear explosion can overcome the gravitational collapse, we first provide a general overview of the different modes we observe in our ECSN simulations.
Here, we focus on a mostly phenomenological description; we present a more in-depth discussion in the Section~\ref{sec:flamespeeds}.
Figure~\ref{fig:expl_modes} illustrates the four different explosion and collapse outcomes.
Here, all models start with the same inital flame, but with different values of \rhoini.

First, we have the prompt explosion, depicted in the first row of Figure~\ref{fig:expl_modes}, showing the example of the \modt{9.95}{73} model.
In this case, the flame starts to rise buoyantly almost immediately, accelerating, and the flame propagation becomes dominated by its turbulent flame speed (see \citealt{jones2016a}).
The gravitational collapse is quickly stopped and the explosion can unbind enough material to leave behind a stable WD remnant. 
Importantly, the remnant is composed of O and Ne at its core, with only a subtle contribution from the burnt ashes in a surface layer.
This is very similar to typical pure deflagrations in CO WDs with a single spot off-center ignition (see, e.g., \citealt{lach2022b}).

Second, at a somewhat higher central density in the \modt{10.11875}{73} model (second row of Figure~\ref{fig:expl_modes}), we find the marginal explosion mode.
In this tECSN mode, the flame just barely manages to halt the gravitational collapse.
As we show later in Section~\ref{sec:flamespeeds}, small changes to, for example, the laminar flame speeds can overturn this outcome.
The distinct difference from the previous prompt explosion mode is that here, the flame (or more precisely the ashes) sinks into the core of the WD, incinerating, and thereby neutronizing the material in the core.
This substantially reduces the effective $\mch$ ($M_\mathrm{Ch,eff}$), further favoring the cECSN outcome.
Here, \rhoini{} is still low enough so that thermonuclear burning prevails and a tECSN is the result.
What further sets this scenario apart from the previous one (and pure deflagrations with a single-spot off-center ignition in CO WDs in general) is that the burning front incinerates the core, leaving behind a bound ONeFe remnant.
In this case, the core of the remnant is to a large fraction composed of burning products, importantly neutron-rich Fe-group elements.
This raises the question whether the remnant will be stable or not, since these ashes will result in a $M_\mathrm{ch,eff}$ value lower than for our initial WD and only very little mass is ejected for some of our models (in extreme cases less than $0.1\,\msun$).
In this work, we do not follow the detailed evolution of the bound remnant.
A more detailed investigation of the fate of the remnant object will be left for future work.

Third, in the \modt{10.146875}{73} model shown in the third row of Figure~\ref{fig:expl_modes}, by just slightly increasing the density, we find the marginal collapse mode.
It is similar to the previous one, with the difference that here the gravitational pull wins and the WD irreversibly collapses.
Importantly, however, the collapse does not manage to completely overpower the burning everywhere (see Section~\ref{sec:flamespeeds} for a description of this mechanism), and part of the flame rises outward and begins to incinerate the outer shell.
Due to the limitations of our code, we cannot follow the collapse, so it remains to be seen if the simulation leads to the formation of a neutron star and if the burning leads to any significant mass ejection.

Last, in our highest density model \modt{10.4}{73}, shown in the last row of Figure~\ref{fig:expl_modes}, we find the prompt collapse mode.
Here, gravity is able to almost immediately overtake the runaway burning.
More precisely, the drop in $Y_e$ due to electron captures happens so fast that the bubble loses too much pressure support without being able to rebuild it efficiently through nuclear burning.
The bubble is then compressed by the surrounding fluid, which has a higher pressure, and its density increases and sinks buoyantly into the WD core.
The neutronization of hot material (burning to an extent continues; it is just no longer dynamically relevant) further destabilizes the WD core so fast that collapse is inevitable.
What appears to be important here is that the ashes sink into the core much faster than in, for example, the marginal collapse case (see the first column of Figure~\ref{fig:expl_modes} where different densities at $0.1\,\mathrm{s}$ are illustrated), causing this crucial region to neutronize faster.
It should be noted that, in all collapsing cases, the thermonuclear burning accelerates the collapse.
This is both due to the increase in temperature and density that increases electron capture rates, as well as the ashes enabling electron captures on, for example, free protons, which is significantly more efficient than electron captures on ${^{20}}\mathrm{Ne}$.
Therefore, while electron captures on ${^{20}}\mathrm{Ne}$ alone are sufficient for a runaway collapse, thermonuclear burning additionally accelerates the collapse if the ashes sink into the core.

In general, we find these four modes for all ignition locations.
However, the r2.5 and r10 series do not exhibit sinking of the ashes toward the core (at least not to the extent that they interact with the flame front) as our ignition hot bubbles already encompass this region due to their size; even in the r20 series, only a marginal amount of sinking ashes can be observed.
How the sinking ashes interact with the flame front critically depends on the laminar flame speed.
We explore this relationship in the following Section~\ref{sec:flamespeeds}.

\section{Explosion versus collapse}\label{sec:flamespeeds}
In this section, we examine the impact of the choice of laminar flame speed parameterization on the outcome of our ECSN simulations and discuss what physical mechanisms determine the collapse or explosion outcome.
Specifically, we compare the laminar flame speed parameterizations of \citetalias{timmes1992a}, given as
\begin{equation}
    v_\mathrm{lam} = 51.8\left(\frac{\rho_9}{6}\right)^{1.06}\left(\frac{X({^{16}}\mathrm{O})}{0.5}\right)^{0.688}\,\mathrm{km}\,\mathrm{s}^{-1},
\end{equation}
to that of \citetalias{schwab2020a}, given as
\begin{equation}
    v_\mathrm{lam} = 16.0\,\rho_9^{0.813}\left[1+96.8\,(0.5 - Y_e)\right]\,\mathrm{km}\,\mathrm{s}^{-1}.
\end{equation}
We note that the laminar flame speeds of \citetalias{schwab2020a} are typically slower than the ones of \citetalias{timmes1992a} for the same fuel composition, see figure~5 of \citetalias{schwab2020a}.
However, as already stated in Section~\ref{sec:methods}, the parameterization of \citetalias{timmes1992a} does not consider cases where $Y_e\neq0.5$, whereas \citetalias{schwab2020a} does.
As illustrated in figure~9 of \citetalias{schwab2020a}, the low $Y_e$ value is the main driver of the large difference in laminar flame speeds.

The total flame speed $v_\mathrm{f}$ is then given by the formula
\begin{equation}
    v_\mathrm{f}=v_\mathrm{lam}\sqrt{1+\frac{4}{3}\left(\frac{v_\mathrm{turb}}{v_\mathrm{lam}}\right)^2},
\end{equation}
following \citet{schmidt2006b}, where $v_\mathrm{turb}^2=2K_\mathrm{sgs}/\rho$ and $K_\mathrm{sgs}$ is the subgrid-scale turbulent kinetic energy.
For turbulent flame speeds $v_\mathrm{turb}$ much higher than the laminar flame speed, the total flame speed asymptotically approaches
\begin{equation}
    v_\mathrm{f}\approx\frac{2 v_\mathrm{turb}}{\sqrt{3}};
\end{equation}
that is, it becomes dominated by the turbulent flame speed.
In contrast $v_\mathrm{f}\approx v_\mathrm{lam}$ if $v_\mathrm{turb} \ll v_\mathrm{lam}$.

\subsection{Neutronization of the core}\label{sec:neutron}
The laminar flame speeds are important during the initial burning before the flame propagation becomes dominated by the turbulent flame speed, at least in the exploding cases.
Although the initial laminar flame growth rate impacts the subsequent nuclear energy release through a larger burning volume at the time when the turbulent propagation becomes dominant, in this section we focus on its impact on the question of whether or not a model collapses or explodes.
For 1D flames, this has already been extensively studied by \citetalias{timmes1992a}, and we can now verify some of their hypotheses based on our 3D simulations.
\begin{figure*}[!hbt]
    \centering
    \includegraphics[height=0.85\textheight,keepaspectratio]{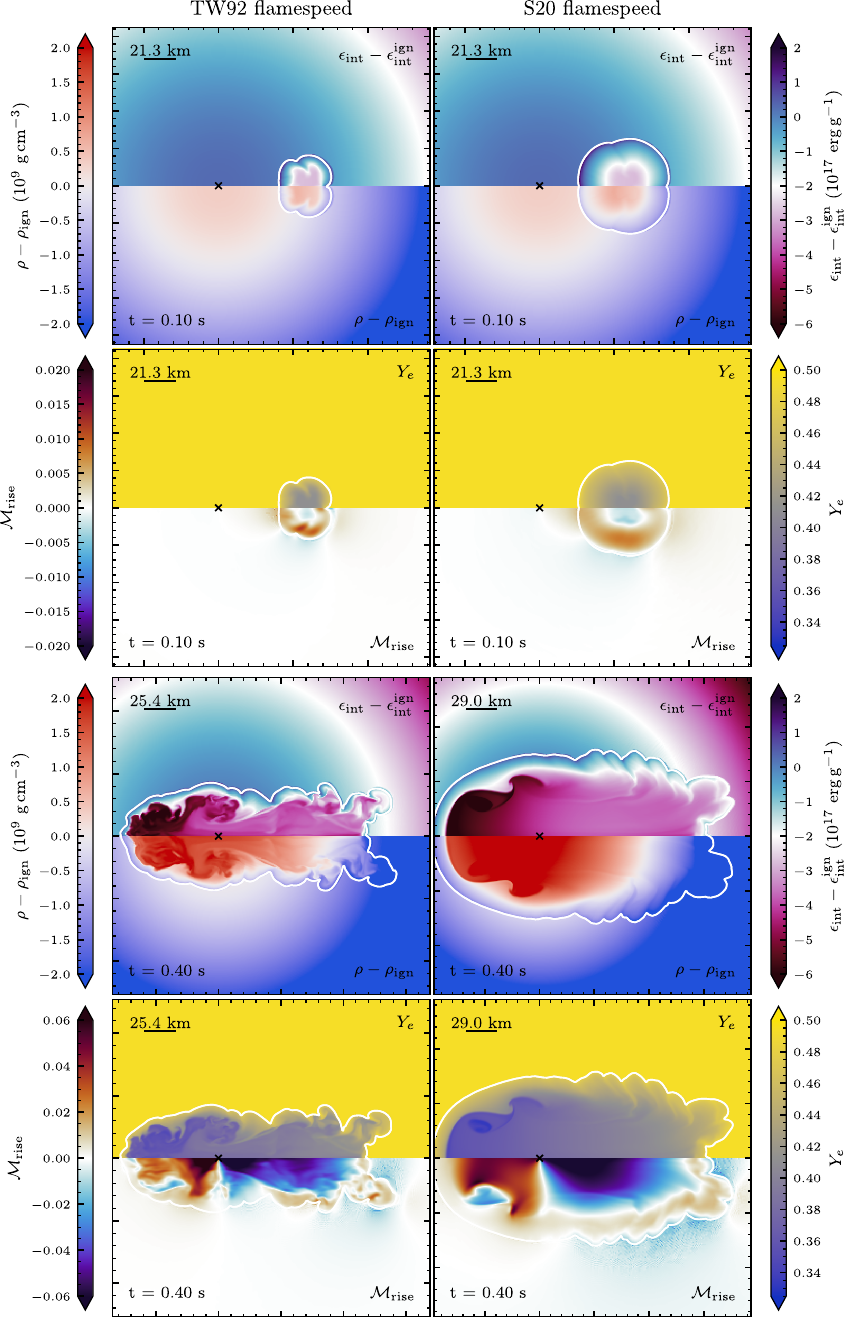}
    \caption{Visualization of the impact of different laminar flame speed parameterizations on the example of the \modt{10.093130625}{50}(\_sfs) model. The left column shows the simulation with the \citetalias{timmes1992a} flame speeds, the right column the one with the \citetalias{schwab2020a} values. The top two rows show the simulation a $t=0.1\,\mathrm{s}$, the bottom two at $t=0.4\,\mathrm{s}$. Here, we show four different quantities: the density and internal energy with respect to their values at the ignition location (at $t=0.0\,\mathrm{s}$), the mach number of the flow in the radial direction $\mathcal{M}_\mathrm{rise}$, and the electron fraction $Y_e$. Note the different ranges of the $\mathcal{M}_\mathrm{rise}$ color bars.}
    \label{fig:flamespeed}
\end{figure*}

\begin{figure}
    \centering
    \includegraphics[]{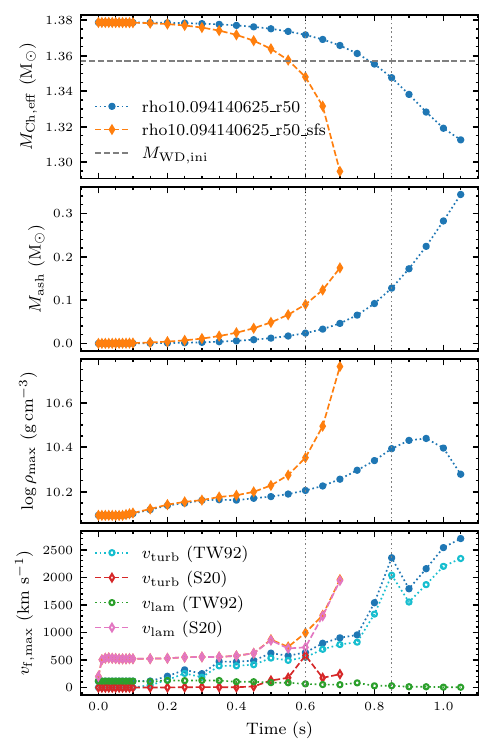}
    \caption{Various quantities related to the neutronization of the core in the comparison between the \citetalias{timmes1992a} and \citetalias{schwab2020a} laminar flame speed parameterizations. First row: Effective $M_\mathrm{Ch}$, computed using the mass weighted value of $Y_e$ over the entire WD. The horizontal dashed line indicates the mass of the initial WD. Second row: Total mass of the ash $M_\mathrm{ash}$, i.e., mass inside the level set. Third row: Maximum density $\rho_\mathrm{max}$ over time. Last row: Maximum total flame speed $v_\mathrm{f,max}$ of the flame anywhere on its surface. Here we also show the values of both the turbulent and laminar flame speed contribution. The vertical dotted lines indicate the time of the first simulation snapshot that where $M_\mathrm{Ch,eff}$ is well below the initial WD mass.}
    \label{fig:quenching}
\end{figure}

In Figure~\ref{fig:flamespeed}, we illustrate the initial stages of the \modt{10.094140625}{50}(\_sfs) model for both laminar flame speed prescriptions.
Importantly, the model variant using the \citetalias{timmes1992a} flame speeds results in a tECSN, whereas the one using the \citetalias{schwab2020a} flame speeds results in a cECSN.
It should be mentioned that in the \modt{10.094140625}{50} model, we also reach the limit of our $\dot{Y}_e$ table, and therefore the outcome is not entirely certain; the arguments we present here are based on the phases before we reach this point.
In the upper half of Figure~\ref{fig:flamespeed}, at $t=0.1\,\mathrm{s}$, it becomes immediately apparent that the higher laminar flame speed in the \citetalias{schwab2020a} case leads, as expected, to the burning of a larger volume.
See also the second row of Figure~\ref{fig:quenching}, which illustrates the total mass of the burned material over time.
This also leads to a greater nuclear energy release, and therefore it is somewhat counterintuitive that this case will collapse because the increase in energy release should provide more stability against collapse.
However, the faster flame speed leads to the neutronization of a larger mass; we remind the reader that electron captures in the ashes are much more efficient, both due to the increase in temperature as well as electron captures on nuclei other than ${^{20}}\mathrm{Ne}$, for example, free protons.
Consequently, $M_\mathrm{Ch,eff}$ drops faster, see the first row of Figure~\ref{fig:quenching}, and the WD begins to contract earlier.
As can be seen in the lower half of Figure~\ref{fig:flamespeed}, at $t=0.4\,\mathrm{s}$, the \mods{10.094140625}{50} model exhibits higher radial Mach numbers $\mathcal{M}_\mathrm{rise}$ toward the center, i.e., the contraction is much more advanced than in the \modt{10.094140625}{50} model.
This is also evident in the $\rho_\mathrm{max}$ values shown in Figure~\ref{fig:quenching}, where the density of the \mods{10.094140625}{50} model increases sharply after $0.4\,\mathrm{s}$ compared to the \modt{10.094140625}{50} case.
At densities examined in our studies, the free-fall time scale is rather short, with $t_\mathrm{ff}\approx0.02\,\mathrm{s}$, and therefore, the evolution of the central density is extremely sensitive to changes in the effective $\mch$.

This contraction leads nuclear burning becoming less efficient.
What happens is that the energy release from burning to NSE increases with density, up to a certain threshold.
Beyond this threshold, the energy release decreases again due to, for example, photodisintegration.
Further contributing to this destabilizing effect is the circumstance that neutronization of the material in the ashes substantially reduces the internal energy; see the upper half of the first row in Figure~\ref{fig:flamespeed}, removing a stabilizing contribution.

This is in large parts due to energy loss from weak neutrinos generated from electron capture reactions, which at the present densities escape freely without interacting with the material.\footnote{Neutrino-matter interaction starts to become relevant at $\rho\geq10^{11}\,\mathrm{g}\,\mathrm{cm}^{-3}$ (see, e.g., \citealt{boccioli2024a,suzuki2024a}). Some of our cECSN simulations nominally reach this value, and it remains to be seen if neutrino-matter interaction plays an important role in the subsequent evolution. \citet{takahashi2019a}, for example, argue that interactions with neutrinos can significantly contribute to the flame propagation speed in this regime.}
\begin{figure}
    \centering
    \includegraphics[]{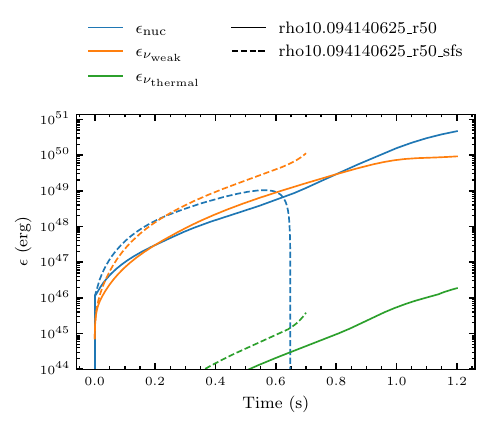}
    \caption{Total (cumulative) energy generated from nuclear reactions over time as well as energy lost from weak and thermal neutrinos.}
    \label{fig:neutrino}
\end{figure}
Critically, as illustrated in Figure~\ref{fig:neutrino}, in both the \citetalias{timmes1992a} and \citetalias{schwab2020a} cases the energy loss through weak neutrinos (i.e., from electron capture reactions) is greater than the energy generated from nuclear burning, with the exception of the initial phase, where the material is not yet rapidly neutronizing over a large volume.
This means that the net energy generation is negative, contributing to the overall contraction of the WD.
In this figure, it can also be seen that the \mods{10.094140625}{50} model exhibits a negative gradient in nuclear energy generation, whereas in the \modt{10.094140625}{50} case the release of nuclear energy substantially increases, eventually leading to a net positive energy release.
At this point, the WD starts to expand again, leading to a slower neutronization and a decrease in energy loss through neutrinos.
In this context, we briefly mention thermal neutrinos as well.
As illustrated in Figure~\ref{fig:neutrino}, their contribution to the overall energy balance is negligible; the largest contribution to the thermal neutrino energy loss comes from plasmon-neutrinos and photo-neutrinos.
It should be noted that this balance is specific to these initial conditions.
In cases where the neutronization is much less prevalent (e.g., the \modt{10.0625}{73}(\_sfs) models), this balance reverses and the nuclear energy generation always outweighs the energy loss from weak neutrinos; in these cases a tECSNe is the clear result anyway.

As collapse (or at least contraction) has already set in, the energy release needs to increase to rebuild the pressure support against gravitational collapse.
Therefore, once the core has contracted beyond this threshold, the burning rate needs to increase to prevent collapse.
Although the rate at which material is burned increases in the \mods{10.094140625}{50} model, see the slope of $M_\mathrm{ash}$ in Figure~\ref{fig:quenching}, it is not enough to rebuild the pressure support and overcome the energy loss from neutrinos and the core collapses.

However, looking at the values of $M_\mathrm{Ch,eff}$, $M_\mathrm{ash}$, and $\rho_\mathrm{max}$ in Figure~\ref{fig:quenching}, both models eventually reach a point where the core becomes unstable and begins to contract; on the surface the difference in laminar flame speed seems to only lead to a relative time difference in contraction occurring.
Similarly, neither model achieves net positive energy generation when considering the energy lost through, for example, weak neutrinos.
Moreover, considering the maximum total flame speed $v_\mathrm{f,max}$, the \mods{10.094140625}{50} seems to exhibit overall faster flame speeds for most of its evolution (which can also be seen at the larger physical extent of the burned region in Figure~\ref{fig:flamespeed}).
Consequently, the question arises why the \mods{10.094140625}{50} model collapses, whereas the \modt{10.094140625}{50} model does not.
The key here is the development of turbulence and its contribution to the maximum flame propagation speed, as we show in the next section.

\subsection{Development of turbulence}\label{sec:turbulence}
\begin{figure}
    \centering
    \includegraphics[width=\linewidth,keepaspectratio]{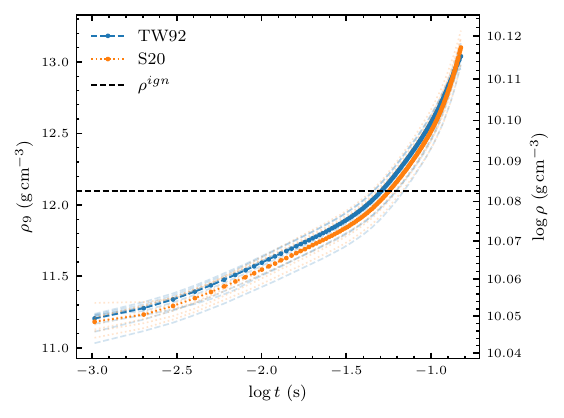}
    \caption{Illustration of the density inversion of the \modt{10.094140625}{50} model for the \citetalias{timmes1992a} and \citetalias{schwab2020a} flame speed parametrization. Here, we show the density trajectories of tracer particles that are located inside the ignition bubble at $t=0.0\,\mathrm{s}$. The transparent lines are the individual trajectories, whereas the opaque lines are the mean density of all trajectories.}
    \label{fig:inversion_dens}
\end{figure}
An important mechanism in tECSNe is, as investigated by \citet{jones2016a}, the development of turbulence and its impact on the flame propagation.
In this section, we investigate the role of turbulence in preventing gravitational collapse.
For a 1D deflagration in ONe WDs (as well as for CO WDs), this has been extensively studied by \citetalias{timmes1992a}, estimating criteria for explosion vs. collapse.
We now reexamine their results for our 3D simulations; importantly, our simulations do not assume a central ignition.

For completeness, we briefly summarize the relations already derived by \citetalias{timmes1992a}.
The material passed by the burning front in scenarios such as ours heats up substantially, causing a decrease in density.
The resultant density gradient, in combination with the gravitational acceleration, causes the formation of a RT instability.
Here, \citetalias{timmes1992a} define the period\footnote{See \citetalias{timmes1992a} for a discussion on defining a period for a nonperiodic motion such as an RT instability} or growth rate as
\begin{equation}
    T^2 = 4\pi\lambda\left(g\frac{\Delta\rho}{\bar{\rho}}\right)^{-1},
\end{equation}
with $\lambda = 2\pi/k$ being the wavelength of the instability and $\Delta\rho=\rho_2-\rho_1$, $\bar{\rho}=(\rho_1+\rho_2)/2$ being the density differential and average across the flame front.
From this quantity, it can be deduced that for any small-scale deformation to form (for a given laminar flame speed), it needs to be larger than a certain minimal scale $\lambda_\mathrm{min}$, which can be written as
\begin{equation}
    \lambda_\mathrm{min} = 4\pi v_\mathrm{lam}^2\left(g\frac{\Delta\rho}{\bar{\rho}}\right)^{-1}.
\end{equation}
Here, the laminar flame speed $v_\mathrm{lam}$ is equivalent to the conductive flame speed $v_\mathrm{cond}$ in the nomenclature of \citetalias{timmes1992a}.

Next, \citetalias{timmes1992a} assume isobaric conditions for the NSE material.
Although we find small-scale deviations from isobaric conditions, we still find it a good enough approximation for the following arguments.
The electron density inside the bubble drops as a result of electron captures, and consequently the pressure drops as dictated by the EOS.
The pressure contrast with the surrounding regions compresses the bubble, thus increasing its gas density and pressure until pressure equilibrium is finally restored (see, e.g., \citealt{cox2006a}).
Because the electron-capture reaction time scale inside the bubble is $\mathbin{\approx}100$ times longer than the sound-crossing time over the size of the bubble, compression occurs subsonically and one can accurately assume that electron captures occur under isobaric conditions.
Therefore, we can define a so-called recovery time $t_\mathrm{recov}$ after which the density has recovered to the value it had before the flame passed the material.

\citetalias{timmes1992a} illustrate this for various initial densities in their figure~8.
We also find this density inversion and recovery time in our simulations.
In both timesteps shown in Figure~\ref{fig:flamespeed} it can clearly be seen that the density immediately behind the flame initially drops but begins to increase to its original value after a while, even increasing beyond this value.
We note that this density increase is the reason for the sinking ashes we observe in many of our simulations.
Initially, when at a lower density, the ashes begin to buoyantly rise (see the $\mathcal{M}_\mathrm{rise}$ values in Figure~\ref{fig:flamespeed}), but electron captures are so fast that the density rapidly increases before the ashes can rise to regions of sufficiently lower densities.
Consequently, the ashes begin to sink to the center of the WD (see the $\mathcal{M}_\mathrm{rise}$ values in Figure~\ref{fig:flamespeed}).
In this context, we point out the relevance of the laminar flame speed.
In the \citetalias{timmes1992a} case, the ashes sink with a velocity comparable to $v_\mathrm{lam}$.
This causes the fluid motion to interact with the flame front, causing its perturbation.
In contrast, the \citetalias{schwab2020a} $v_\mathrm{lam}$ is much larger (as discussed in the introductory paragraph of Section~\ref{sec:flamespeeds} this is due to the \citetalias{schwab2020a} parameterization considering the impact of decreasing $Y_e$), and therefore, the ashes do not interact with the flame front (at least not to the same degree).
This is evident in Figure~\ref{fig:flamespeed}, where the much smoother flame front can be seen in the \citetalias{schwab2020a} case (in the center).
Furthermore, the RT instabilities in the ashes are much less pronounced than in the \citetalias{timmes1992a} case.

Returning to the recovery time $t_\mathrm{recov}$ described by \citetalias{timmes1992a}, we briefly analyze the value found for our \modt{10.094140625}{50}(\_sfs) models.
For this, we follow the density evolution of the tracer particles that are inside the initial ignition bubble at $t=0.0\,\mathrm{s}$.
The result is illustrated in Figure~\ref{fig:inversion_dens}.
Here, we show both the individual tracer trajectories and the average evolution.
The horizontal line indicates the initial density at the ignition location.
From this, $t_\mathrm{recov}$ is immediately apparent.
As expected, the resultant $t_\mathrm{recov}\approx0.05\,\mathrm{s}$ is independent of the choice of $v_\mathrm{lam}$ as it is connected to the electron capture process; the slight differences between the two flame speed prescriptions in Figure~\ref{fig:inversion_dens} is due to the stochasticity of the tracer particles.
It is rather reassuring that we obtain a value quite similar to \citetalias{timmes1992a}, who find $t_\mathrm{recov}\approx0.03\,\mathrm{s}$.
Given the different setups (density and composition) as well as the included physics, a small difference is expected.

From the existence of this recovery time, \citetalias{timmes1992a} infer a finite thickness of the density inversion region and from this an upper limit on the length scale of deformations:
\begin{equation}
    \lambda_\mathrm{max} = v_\mathrm{lam}\,t_\mathrm{recov}.
\end{equation}
As a consequence, once $\lambda_\mathrm{min} > \lambda_\mathrm{max}$, no deformations can form and the flame will propagate in a stable manner.
\citetalias{timmes1992a} suspect this crossover as the deciding factor between explosion and collapse, that is, if $\lambda_\mathrm{max} > \lambda_\mathrm{min}$ the outcome is a thermonuclear explosion and vice versa, see their figure~9b.
While their analysis exclusively focuses on the conductive propagation of the flame, our simulations also consider the contribution of turbulence to the propagation speed of the flame, and therefore we have the opportunity to revisit their results in a more complete setup.
Looking at their figure~9b, we find that at $\rhoini = 12.4\times10^9\,\mathrm{g}\,\mathrm{cm}^3$ of our \modt{10.094140625}{50} model they predict a thermonuclear explosion by a small margin.
Note that their case A, which we compare here has a lower ${^{16}}\mathrm{O}$ mass fraction than our model, but for this qualitative argument we deem this a close enough match.
Their prediction is confirmed by our simulation, and our \modt{10.094140625}{50} model results in a thermonuclear explosion using the prescription of \citetalias{timmes1992a}.

\begin{figure*}
    \centering
    \includegraphics[]{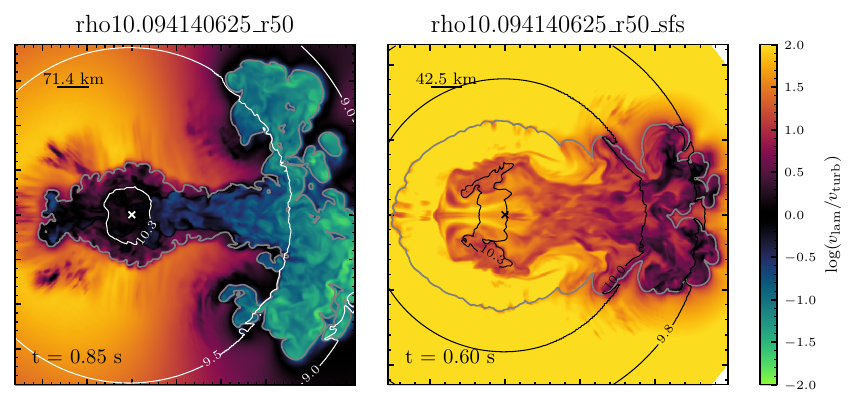}
    \caption{Illustration of the ratio between $v_\mathrm{lam}$ and $v_\mathrm{turb}$. We show the \citetalias{timmes1992a} (left) and \citetalias{schwab2020a} (right) variants of the\modt{10.094140625}{50} model at the point in time where their $M_\mathrm{Ch,eff}$ drops below the initial WD mass. The gray contour indicates the levelset, whereas the white and black contours indicate surfaces of equal $\log\rho$. We note that although we show $v_\mathrm{lam}/v_\mathrm{turb}$ on the whole grid, it is only relevant at the flame surface; we only show it everywhere else for illustration purposes.}
    \label{fig:lamoturb}
\end{figure*}
In contrast, our simulations predict a collapse for the same model when using the \citetalias{schwab2020a} parameterization.
This is particularly surprising because, as discussed in the previous Section~\ref{sec:neutron}, a faster flame speed should lead to a larger nuclear energy release, thereby providing more energy to counteract the collapse.
Or, looking at this the other way around, both models generate less nuclear energy than is lost through neutrinos during the contraction phase, and therefore one could expect that both models collapse.
Additionally, the simulation using the \citetalias{timmes1992a} parameterization shows the same contraction after accounting for the relative time difference in the neutronization of the core due to the slower flame speed.
Therefore, the question arises as to why one model collapses, whereas the other does not.

The answer lies in the development of turbulence and its impact on the flame propagation.
While in the model using the \citetalias{timmes1992a} parameterization the flame propagation becomes dominated by its turbulent contribution rather quickly (see Figure~\ref{fig:quenching}), the faster $v_\mathrm{lam}$ of \citetalias{schwab2020a} suppresses the formation of instabilities.
Essentially, because $\lambda_\mathrm{max}\propto v_\mathrm{lam}$ but $\lambda_\mathrm{min}\propto v_\mathrm{lam}^2$, the minimum unstable length scale will grow faster with $v_\mathrm{lam}$ than the maximum deformation length scale $\lambda_\mathrm{max}$ and the flame will remain stable.

The turbulent propagation of the flame is important as it can thereby reach much higher speeds than its laminar counterpart.\footnote{Figure~\ref{fig:quenching} suggests that the \mods{10.094140625}{50} model has a much larger total flame speed than the \modt{10.094140625}{50} model for $t\lesssim0.7\,\mathrm{s}$. However, the values shown here originate from the dense and neutron rich core region, not the rising part of the flame. In those regions, the difference between the two models ($v_\mathrm{lam}$ in particular) is much smaller.}
This is critical as this allows the flame to burn much farther outward before the core region becomes unstable due to its neutronization.
Since the flame thereby reaches regions of lower density, it is less susceptible to the decrease in efficiency described in Section~\ref{sec:neutron}.
The flame can thus effectively rebuild the pressure support needed to prevent collapse and lead to an explosion, as is evident in the decreasing $\rho_\mathrm{max}$ in Figure~\ref{fig:quenching}.
Figure~\ref{fig:lamoturb} illustrates this mechanism, where we show both models at the point in time where their $M_\mathrm{Ch,eff}$ drops below the initial WD mass.
Here, it can be seen that in the \mods{10.094140625}{50} case, although in the rising part of the flame $v_\mathrm{turb}$ becomes comparable to $v_\mathrm{lam}$, the flame is still mostly propagating with its laminar speed.
Importantly, burning occurs at comparably high densities where it is susceptible to becoming less efficient.
In contrast, the rising part of the flame in the \modt{10.094140625}{50} model is almost completely dominated by $v_\mathrm{turb}$ and burns at significantly lower densities.
It should be noted that at the respective simulation time, both models have reached a similar central density due to the initial phase of contraction.
However, the flame in the \modt{10.094140625}{50} case has burned much farther outward (see the scale indicators in Figure~\ref{fig:lamoturb}), where it is able to halt the collapse.

In conclusion, the formation of turbulence does not prevent the collapse by itself.
Instead, it allows the flame to propagate fast enough to reach regions of lower density where burning is efficient enough to prevent the collapse.
At these lower densities electron captures are also slower, thereby reducing the energy loss through neutrinos; see the reduction in energy loss in Figure~\ref{fig:neutrino}.
Although we content ourselves with only a qualitative analysis here, it is nevertheless satisfying to see that the hypothesis of \citetalias{timmes1992a} appears to be confirmed by modern 3D simulations.

We note that in our simulations, flame propagation starts fully laminar as turbulence has yet to develop.
However, the flame could already be dominated by turbulence by the time it reaches scales resolved by our numerical grid.
Here, the arguments made above apply as well and it depends on factors such as $v_\mathrm{lam}$, $t_\mathrm{recov}$, and the density profile if turbulence is suppressed or if instabilities can form.
Therefore, this is less of a concern for our higher density models, for example, the \mods{10.094140625}{50} model discussed above, where instabilities are suppressed.
In contrast, models such as the \modt{9.95}{73} may already be fully dominated by the turbulent flame propagation by the time the flame reaches scales resolved by our computational grid, and one has to distinguish between model and reality.

\section{Parameter study}\label{sec:paramstudy}
\begin{figure*}
    \centering
    \includegraphics[width=\linewidth,keepaspectratio]{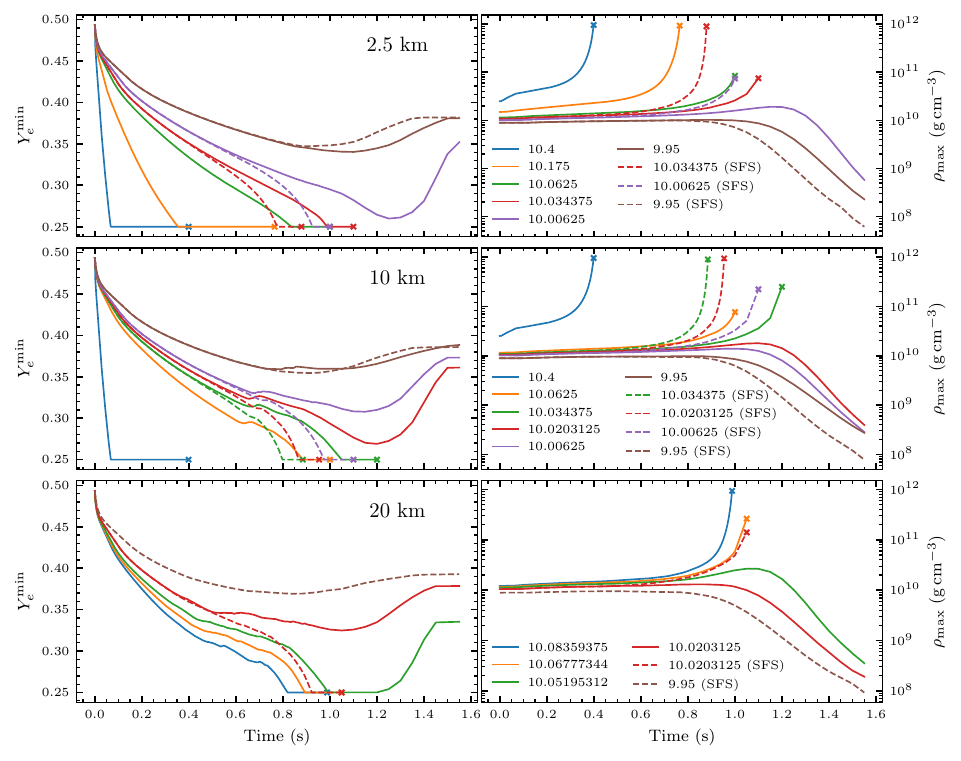}
    \caption{Simulation history of all simulations in our parameter study. The left column shows the minimal electron fraction $Y_e^\mathrm{min}$, and the right column shows the maximum density $\rho_\mathrm{max}$. Both quantities are the extrema found anywhere in the simulation. These trajectories are used to estimate if a simulation results in an explosion or a collapse. (Continued in Figure~\ref{fig:protocol2}.)}
    \label{fig:protocol1}
\end{figure*}
We now focus on the results of our parameter study, that is, for which \rhoini{} and $r_\mathrm{ign}$ combinations our simulations predict a thermonuclear explosion or a collapse.
Before discussing our results, we briefly remind the reader of what we interpret as a thermonuclear explosion versus a collapse outcome.
Since our \textsc{Leafs} code cannot follow the full collapse to, for example, an NS, rather only the onset of collapse, and our code is limited to $\ye\geq0.25$, we can only give an initial prediction if a simulation actually collapses.
These predictions will require confirmation in full core-collapse simulations.
Therefore, we defined a simulation to predict a cECSNe if we observe a large volume reaching $\ye=0.25$ for an extended period of time and a rapid increase of $\rho_\mathrm{max}$ by several orders of magnitude.
\citet{jones2016a} stopped following simulations once they reach $\ye=0.25$, but for their simulations this was only a necessary condition for a collapse; they also identified the collapse based on a rapid density increase.\footnote{On a more qualitative level, it is usually quite apparent when a simulation will collapse by looking at a visualization, where the collapse can be clearly seen.}
For the density extremes (low and high) in our study, the outcome is usually clearly identifiable.
However, once we approach the transition regime, it is less clear how to interpret our simulations, in particular, when simulations start to reach the edges of our $\dot{\ye}$ table.
As long as only a few cells reach this limit, it will most likely not affect the outcome, and we tolerate $\ye=0.25$ in our explosion vs. collapse decision.
Since we do not observe cases where $\ye=0.25$ is reached in a large volume for an extended period of time and simultaneously see a tECSNe outcome, we deem this acceptable.
In short, we only classify simulations as cECSN if we observe a rapid increase in density of several orders of magnitude.

Figures~\ref{fig:protocol1} and \ref{fig:protocol2} illustrate the resultant $\ye^\mathrm{min}$ and $\rho_\mathrm{max}$ trajectories of all our simulations, both for simulations using the \citetalias{timmes1992a} and \citetalias{schwab2020a} flame speed prescriptions.
Here, we mark the models that we interpret as cECSNe with ``x'' at the end of their trajectories.
From this, it can be seen that collapsing models experience a sharp density increase, whereas exploding models show, if at all, only a modest increase in $\rho_\mathrm{max}$ before the explosion unbinds substantial amounts of material.
As explained in Section~\ref{sec:flamespeeds}, this increase in density comes from an initial phase of contraction until burning can become efficient enough to outweigh, for example, the energy loss from neutrinos.
From these figures, it can also be seen that for some of the more marginal cases, such as \modt{10.05195312}{20} or \modt{10.11875}{73} the WD initially, over $\approx1.0\,\mathrm{s}$, the model contracts with $\rho_\mathrm{max}$ gradually increasing.
In these cases, $\ye^\mathrm{min}$ decreases to $\ye=0.25$ before contraction stops.
However, this is only in a small region in the central part of the WD, where the ashes have sunk into the core.
In the outer region, the flame is growing steadily until it can eventually unbind enough material to sufficiently reduce the pressure of the core; at this point $\rho_\mathrm{max}$ decreases again.

The limitations of our $\dot{\ye}$ table can also be seen, for example in the \modt{10.4}{73} model.
Here $\ye^\mathrm{min}=0.25$ is reached within $\approx0.05\,\mathrm{s}$.
Looking at the corresponding $\rho_\mathrm{max}$ trajectory in Figure~\ref{fig:protocol2}, one can observe a kink and the density increase somewhat slows down.
This is, of course, unphysical.
The density increase caused by a reduction in \ye\ no longer contributes, and therefore $\rho_\mathrm{max}$ increases more slowly.
However, in these simulations the outcome is not affected by this effect.
By the time $\ye^\mathrm{min}=0.25$ is reached, the WD core is already starting to collapse and the burning effectively only contributes by its neutronizing effect rather than its nuclear energy release (see Section~\ref{sec:neutron}), and we can confidently predict that these models will end up as a cECSN.

Looking closer at some of the \yemin\ trajectories, for example, of \modt{10.08359375}{50} in the second row of Figure~\ref{fig:protocol2}, we can see two characteristic features.
The first is the rapid drop in \yemin\ for $t>0.8\,\mathrm{s}$.
In the beginning \yemin\ drops comparably slow, but after this point the neutronization seems to increase rapidly.
This is because for $t<0.8\,\mathrm{s}$, the electron captures occur approximately at a constant density.
However, as the ashes sink into the core, they destabilize the WD and it begins to contract.
This causes the electron capture rates in the core to increase (see also the $\rho_\mathrm{max}$ trajectory).
In these models, the outward burning flame manages to trigger an explosion just in time before the WD can collapse.
The second feature is that after the explosion is able to unbind the material and the density decreases, the ashes releptonize until the NSE composition freezes out; the \yemin\ trajectory forms a plateau after this point.

We note that, as already discussed by \citet{jones2019a}, our \yemin{} evolution is notably different from previous work on ECSNe.
Specifically, comparing our work to \citet{miyaji1987a}, who find a $\yemin\approx0.34$ at the onset of collapse, whereas we find $\yemin=0.25$.
There is some ambiguity as to how the quantities were computed in both works, but the trend is clear.
As illustrated in \citet[figure 4]{jones2019a}, the inclusion of the electron capture rates by \citet{nabi2004a} leads to a much more rapid drop in $Y_e$ for $Y_e\lesssim0.45$ compared to the case without these rates.
Interestingly, if one assumes the $Y_e$ evolution without the rates by \citet{nabi2004a} several of our cECSN models would likely end up as a tECSN instead.
Comparing our \yemin{} and $\rho_\mathrm{max}$ trajectories to those found by \citet[figure 3]{leung2020a}, we find good qualitative agreement, at least in cases where our simulations agree on the outcome.
Specifically our exploding models achieve a similar \yemin{} as their exploding models for the same \rhoini{}.
This is insofar not surprising as both their and our simulation heavily rely on the electron capture rates of \citet{nabi2004a} for this regime.

\begin{figure}
    \centering
    \includegraphics[width=\linewidth]{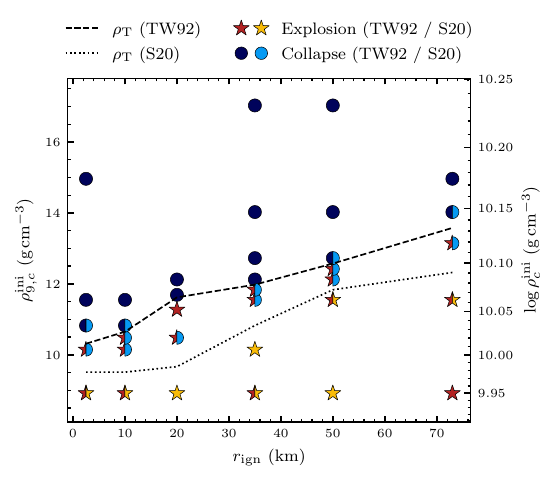}
    \caption{Outcomes of all simulations in our parameter study dependent on ignition location and central density at ignition. Here the dashed and dotted lines indicate the transition from explosion to collapse for the \citetalias{timmes1992a} and \citetalias{schwab2020a} flame speeds, respectively. These lines are only serve as a guide for the eye as they only show the midpoint between the last exploding and first collapsing simulation. We note that this figure does not include simulations with $\log\rhoini=10.4$ for better visibility.}
    \label{fig:expl_vs_collapse}
\end{figure}
In summary of our results, we illustrate the outcome for each simulation we performed in Figure~\ref{fig:expl_vs_collapse}.
The first thing that stands out is that the an off-center ignition generally allows a higher \rhoini, while still resulting in a tECSN.
We emphasize that the same trend is visible for the density at the ignition location, i.e., the transition from explosion to collapse does not seem to occur simply at a fixed ignition density.
As described in Section~\ref{sec:flamespeeds}, the sinking ashes that destabilize the core play a critical role in accelerating the collapse.
Consequently, the time it takes for the ashes to sink (or rise to lower density regions), or in other words, the distance of the ignition from the center, affects the outcome, leading to the trend seen in Figure~\ref{fig:expl_vs_collapse}.
Somewhat related to this is the noticeable jump between the r10 and r20 series.
To an extent, this is over-exaggerated by the uneven resolution of our model grid, but there is also a physical argument to be made.
In the r2.5 and r10 series, the ignition spot covers the WD center at $t=0.0\,\mathrm{s}$ (the width of the ignition bubble is $18.5\,\mathrm{km}$), and the flame immediately starts neutronizing the core.
Importantly, the sinking ashes do not significantly interact with the flame front.

In contrast, the r20 ignition spot initially does not cover the center.
Therefore, some time passes before the ashes start to sink into the core (in addition to the time it takes for the density to increase again).
This gives the flame some more time to grow and potentially result in a tECSN at slightly higher \rhoini.
It remains to be seen how sensitive this is to smaller (or larger) ignition bubbles, but an argument is to be made that for smaller ignition bubbles the laminar flame speed is fast enough to grow the burned area to our nominal ignition bubble size before neutronization can noticeably impact $M_\mathrm{ch,eff}$.

This picture does not change substantially for models that use \citetalias{schwab2020a} flame speeds.
As detailed in Section~\ref{sec:flamespeeds}, the faster flame speed also leads to an accelerated decrease in $M_\mathrm{ch,eff}$ and a more efficient suppression of turbulence.
Therefore, a tECSN seems to only be possible for a comparably lower \rhoini.
However, a thermonuclear explosion is possible for all $r_\mathrm{ignition}$, irrespective of the flame speed parametrization, as long as \rhoini{} is low enough.
Overall, our results agree well with the outcomes predicted by \citet{jones2016a,jones2019a}, even though we use a single-spot ignition instead of a multi-spot ignition (as well as several more technical changes, such as the change to the EOS and the gravity solver, see Section~\ref{sec:methods}).
Moreover, our results are in line with the trend already suggested by \citet{leung2020a} (see their Table 2), i.e., that a more off-center ignition supports the tECSN outcome for higher \rhoini{}.
However, their numerical methods, for example the employed subgrid scale model for turbulence, differ from those implemented in \textsc{Leafs}.
Additionally, their simulations are limited to lower resolution in 2D and only cover one quadrant. 
Therefore, it is difficult to interpret their somewhat lower transition density for a given ignition geometry.

Our parameter study does not consider the impact of varying initial composition and instead keeps the composition of our initial WD fixed at $65\%/35\%$ ONe. 
This values is strongly linked to the evolutionary path required to reach the density regime considered here \citep{jones2013a}.
For example, the presence of residual C can cause an ignition at lower densities during electron capture on ${^{24}}\mathrm{Mg}$ and ${^{25}}\mathrm{Mg}$ \citep{antoniadis2020a}.
This is particularly interesting as the ${^{12}}\mathrm{C}(\alpha,\gamma){^{16}}\mathrm{O}$ reaction rate is especially uncertain \citep[e.g.,][]{an2015a,pepper2022a}, and therefore it is not clear which conditions will actually be reached at O ignition.
However, we estimate our results to be robust against smaller changes to this initial composition, at least relative to the difference in transition densities caused by the choice of laminar flame speed parameterization.
Changes to the O to Ne ratio will, on the one hand, as shown by \citetalias{schwab2020a}, change the laminar flame speed only by a small amount, particularly compared to the difference caused by varying $Y_e$.
On the other hand, a change in initial composition will impact the resulting energy release from nuclear burning, equally affecting the ECSN outcome.
In general, it is not clear in which direction the transition density will shift with an increase in, for example, the Ne fraction.
An increase in the Ne fraction will decrease the nuclear energy release from burning (favoring the cECSN outcome), but simultaneously also decrease the laminar flame speed, thereby reducing the suppression of turbulence (favoring the tECSN outcome).
This also extends to effects such as Ne depletion in the core, that is a non-uniform initial composition, or the contamination by, for example, residual Mg (which can also be non-uniformly distributed).
The likely most relevant effect of the initial composition will be the $Y_e$ reached at ignition, given its notable impact on the laminar flame speed (at least in the \citetalias{schwab2020a} formulation), with higher $Y_e$ values favoring the tECSN outcome.
A nonuniform $Y_e$ composition will be particularly interesting as the suppression of turbulence will be highly localized due to the strong $Y_e$ dependence of the \citetalias{schwab2020a} laminar flame speed.
Further numerical simulations are required to establish to what extent the transition density depends on the initial composition.

\section{Conclusions}\label{sec:conclusion}
In this work, we presented a parameter study of 3D hydrodynamic simulations of ECSNe starting at the time of thermonuclear ignition.
With this study, we investigated under which conditions a deflagration in a high-density ONe core leads to a cECSN or a tECSN, and what physical mechanisms are relevant in determining the outcome.
In total, we conducted 56 full 3D hydrodynamic simulations using the \textsc{Leafs} code.

Here, in Section~\ref{sec:ecsn_modes}, we found four distinct modes in which ECSNe can proceed: the prompt and marginal explosion and collapse cases.
Each of these modes is interesting in its own right, but the marginal cases are of particular interest in the present work.
Not only are these models indicative of the transition from explosion to collapse, but these marginal cases also exhibit (in the case of an off-center ignition) a sinking of the ashes toward the WD core; an effect not seen in deflagrations in CO WDs or lower-density ONe WDs, where the ashes purely rise buoyantly (see, e.g., \citealt{seitenzahl2013a,fink2014a,lach2022b}).
Although the physical effect causing this sinking -- the density inversion due to rapid electron captures \citep{timmes1992a} -- has long been known, only now in full 3D simulations can we observe its impact on the outcome of pure deflagrations.
In particular, the most direct effect of this process is that it accelerates the destabilization of the WD by poisoning its core with neutron-rich material.
Moreover, this means that much of the neutron-rich ashes become trapped in the core.
The resulting bound remnant in the case of a thermonuclear explosion consequentially contains an extremely metal-rich core, and it is not clear if the bound remnant will remain stable for longer periods of time.
An investigation of the bound remnants will be the focus of future research.
Furthermore, it remains to be seen if, in the collapsing cases, the asymmetries introduced by thermonuclear burning and the sinking of the ashes have a significant impact on the subsequent core-collapse.
This is of particular interest for the marginal collapse case, where additionally the rising part of the thermonuclear flame could potentially unbind substantial amounts of material, thereby favoring the formation of extremely low-mass NSs.

In Section~\ref{sec:flamespeeds}, we also investigated the impact of various parameterizations of the laminar flame speed and how the development of turbulence affects the outcome of an ECSN.
Here, we found that, counterintuitively, a higher laminar flame speed favors the collapse outcome.
In general, a faster laminar flame speed leads to higher nuclear energy release.
However, the neutronization of the ashes that sink into the center leads to a contraction of the WD, further raising its density.
Additionally, during this initial phase, nuclear burning is not able to outweigh the energy loss from escaping weak neutrinos, contributing to the contraction.
The contraction and the subsequent increase in density cause nuclear burning to become less efficient (at least for the density ranges found in marginal and collapsing cases).
This process occurs to some extent regardless of the choice of laminar flame speed, but it may be less important in lower density scenarios where electron captures are comparably slow.
However, in the case of the slower flame speeds of \citetalias{timmes1992a}, instabilities can form and flame propagation becomes dominated by the substantially faster turbulent flame speed.
As a consequence, this model can reach regions with lower densities where nuclear burning is efficient again, ultimately making it more likely to produce a tECSN.
In the case of the \citetalias{schwab2020a} flame speeds, turbulence is suppressed and the flame remains laminar and remains confined to high density regions.
Eventually, the burning is overpowered, and the result is a cECSN.
This comparison illustrates the importance of the conductive flame speed in these marginal cases, as it largely determines the outcome.
In the prompt explosion mode (and for similar reasons the prompt collapse), the conductive flame speed is not as relevant as the flame will be rapidly dominated by its turbulent flame speed (see \citealt{jones2016a}) which will then determine most of the subsequent dynamics.

The main result of our parameter study is described in Section~\ref{sec:paramstudy} and illustrated in Figure~\ref{fig:expl_vs_collapse} showing that, depending on the ignition location, a wide range of \rhoini{} supports a tECSN.
Therefore, it is important for stellar evolution models to accurately determine both the location (or locations) at which the thermonuclear burning starts and the central density that the ONe core has reached at this point to determine whether or not the resulting event is a tECSNe or cECSNe.
Although, at the point of writing, recent studies have suggested a range of ignition conditions are possible (see, e.g., \citealt{kirsebom2019a}), an accurate determination of the ignition process will require multidimensional simulations to fully capture the impact of convection and other inherently multidimensional effects (see, e.g., \citealt{denissenkov2013a,jones2016a}).
Moreover, as investigated by \citet{kirsebom2019a}, the ignition process itself is sensitive to specific decay rates (they find that an increased electron capture rate on ${^{20}}\mathrm{Ne}$ leads to an ignition at lower central densities, thereby favoring the tECSN outcome) during the evolution leading up to the runaway.
Therefore, it remains an open question what conditions in terms of \rhoini{} can be reached before the ignition from a stellar evolution perspective.

In summary, we find that tECSNe from high-density ONe WDs are predicted from our simulations, but the outcome strongly depends on the conditions at ignition.
Factors such as laminar flame speed, neutronization rate, nuclear energy release, and turbulence formation play an important role in determining the outcome, particularly in the transition region from explosion to collapse.
Deflagrations in high-density ONe WDs can cause part of the ashes to sink into the WD core, a mechanism not observed in low-density deflagrations, for example, in CO WDs.
This mechanism can potentially have a significant impact on the evolution of the bound remnant object and may alter the ejecta composition by trapping the most neutron rich species in the remnant object.
Additionally, the asymmetric explosion can impart a large kick velocity to the bound remnant object.
Therefore, it remains to be seen whether tECNSe can potentially explain objects such as hyper-velocity LP40-365 stars \citep[e.g.,][]{vennes2017a,raddi2019a,el-badry2023a}, which have been hypothesized to be the remnants of deflagrations in ONe WDs.
More work is needed to determine the long-term fate of these iron-rich bound remnants, as well as the structure and composition of the ejecta and will be the focus of future research.

\begin{acknowledgements}
A.H. and C.F. are fellows of the International Max Planck Research School for Astronomy and Cosmic Physics at the University of Heidelberg (IMPRS-HD) and acknowledges financial support from IMPRS-HD.

This work received support from the European Research Council (ERC) under the European Union’s Horizon 2020 research and innovation programme under grant agreement No.\ 759253 and 945806, the Klaus Tschira Foundation, and the High Performance and Cloud Computing Group at the Zentrum f{\"u}r Datenverarbeitung of the University of T{\"u}bingen, the state of Baden-W{\"u}rttemberg through bwHPC and the German Research Foundation (DFG) through grant no INST 37/935-1 FUGG.
The authors gratefully acknowledge the Gauss Centre for Supercomputing e.V. (www.gauss-centre.eu) for funding this project by providing computing time through the John von Neumann Institute for Computing (NIC) on the GCS Supercomputer JUWELS at Jülich Supercomputing Centre (JSC).
The authors acknowledge support by the state of Baden-Württemberg through bwHPC
and the German Research Foundation (DFG) through grant INST 35/1597-1 FUGG.
This work was supported by the Deutsche Forschungsgemeinschaft (DFG, German Research Foundation) -- RO 3676/7-1, project number 537700965,
and by the European Union (ERC, ExCEED, project number 101096243). Views and opinions expressed are, however, those of the authors only and do not necessarily reflect those of the European Union or the European Research Council Executive Agency. Neither the European Union nor the granting authority can be held responsible for them.
This work was performed on the HoreKa supercomputer funded by the Ministry of Science, Research and the Arts Baden-Württemberg and by the Federal Ministry of Education and Research.
This work was supported by the U.S. Department of Energy through the Los Alamos National Laboratory. Los Alamos National Laboratory is operated by Triad National Security, LLC, for the National Nuclear Security Administration of U.S. Department of Energy (Contract No. 89233218CNA000001).
\end{acknowledgements}

\bibliographystyle{aa}
\bibliography{references}

@article{antoniadis2020a,
  title = {Type {{Ia}} Supernovae from Non-Accreting Progenitors},
  author = {Antoniadis, J. and Chanlaridis, S. and Gr{\"a}fener, G. and Langer, N.},
  year = 2020,
  month = mar,
  journal = {A\&A},
  volume = {635},
  pages = {A72},
  issn = {0004-6361},
  doi = {10.1051/0004-6361/201936991},
  urldate = {2024-06-27}
}

@article{arcones2010a,
  title = {Electron Fraction Constraints Based on Nuclear Statistical Equilibrium with Beta Equilibrium},
  author = {Arcones, A. and {Mart{\'i}nez-Pinedo}, G. and Roberts, L. F. and Woosley, S. E.},
  year = 2010,
  month = nov,
  journal = {A\&A},
  volume = {522},
  pages = {A25},
  issn = {0004-6361, 1432-0746},
  doi = {10.1051/0004-6361/201014276},
  urldate = {2025-08-25},
  copyright = {{\copyright} ESO, 2010},
  langid = {english}
}

@article{badwaik2020a,
  title = {Single-{{Step Arbitrary Lagrangian}}--{{Eulerian Discontinuous Galerkin Method}} for 1-{{D Euler Equations}}},
  author = {Badwaik, Jayesh and Chandrashekar, Praveen and Klingenberg, Christian},
  year = 2020,
  month = dec,
  journal = {Commun. Appl. Math. Comput.},
  volume = {2},
  number = {4},
  pages = {541--579},
  issn = {2661-8893},
  doi = {10.1007/s42967-019-00054-5},
  urldate = {2024-11-27},
  langid = {english}
}

@article{boccioli2024a,
  title = {The {{Physics}} of {{Core-Collapse Supernovae}}: {{Explosion Mechanism}} and {{Explosive Nucleosynthesis}}},
  shorttitle = {The {{Physics}} of {{Core-Collapse Supernovae}}},
  author = {Boccioli, Luca and Roberti, Lorenzo},
  year = 2024,
  month = mar,
  journal = {Universe},
  volume = {10},
  number = {3},
  pages = {148},
  issn = {2218-1997},
  doi = {10.3390/universe10030148},
  urldate = {2025-10-17},
  copyright = {http://creativecommons.org/licenses/by/3.0/},
  langid = {english}
}

@article{butterworth1930a,
  title = {On the {{Theory}} of {{Filter Amplifiers}}},
  author = {Butterworth, Stephen},
  year = 1930,
  pages = {536--541},
  urldate = {2024-04-05}
}

@article{canal1992a,
  title = {The {{Quasi-static Evolution}} of {{ONeMg Cores}}: {{Explosive Ignition Densities}} and the {{Collapse}}/{{Explosion Alternative}}},
  shorttitle = {The {{Quasi-static Evolution}} of {{ONeMg Cores}}},
  author = {Canal, Ramon and Isern, Jordi and Labay, Javier},
  year = 1992,
  month = oct,
  journal = {ApJ},
  volume = {398},
  pages = {L49},
  issn = {0004-637X},
  doi = {10.1086/186574},
  urldate = {2025-10-07}
}

@article{chanlaridis2022a,
  title = {Thermonuclear and Electron-Capture Supernovae from Stripped-Envelope Stars},
  author = {Chanlaridis, S. and Antoniadis, J. and {Aguilera-Dena}, D. R. and Gr{\"a}fener, G. and Langer, N. and Stergioulas, N.},
  year = 2022,
  month = dec,
  journal = {A\&A},
  volume = {668},
  pages = {A106},
  issn = {0004-6361, 1432-0746},
  doi = {10.1051/0004-6361/202243035},
  urldate = {2024-06-10},
  copyright = {https://creativecommons.org/licenses/by/4.0}
}

@article{colella1984a,
  title = {The {{Piecewise Parabolic Method}} ({{PPM}}) for {{Gas-Dynamical Simulations}}},
  author = {Colella, P. and Woodward, Paul R.},
  year = 1984,
  month = sep,
  journal = {J. Comp. Phys.},
  volume = {54},
  pages = {174--201},
  issn = {0021-9991},
  doi = {10.1016/0021-9991(84)90143-8},
  urldate = {2024-03-11}
}

@book{cox2006a,
  title = {Cox and {{Giuli}}'s {{Principles}} of Stellar Structure},
  editor = {Cox, John P. and Weiss, A. and Weiss, Achim and Giuli, R. Thomas},
  year = 2006,
  series = {Advances in Astronomy and Astrophysics},
  edition = {Extended 2. ed., Repr},
  number = {8},
  publisher = {Cambridge Scientific Publishers},
  address = {Cambridge},
  isbn = {978-1-904868-20-0 978-1-904868-55-2},
  langid = {english}
}

@article{denissenkov2013a,
  title = {The {{C-flame Quenching}} by {{Convective Boundary Mixing}} in {{Super-AGB Stars}} and the {{Formation}} of {{Hybrid C}}/{{O}}/{{Ne White Dwarfs}} and {{SN Progenitors}}},
  author = {Denissenkov, P. A. and Herwig, F. and Truran, J. W. and Paxton, B.},
  year = 2013,
  month = jul,
  journal = {ApJ},
  volume = {772},
  pages = {37},
  issn = {0004-637X},
  doi = {10.1088/0004-637X/772/1/37},
  urldate = {2025-09-03}
}

@article{doherty2017a,
  title = {Super-{{AGB Stars}} and Their {{Role}} as {{Electron Capture Supernova Progenitors}}},
  author = {Doherty, Carolyn L. and {Gil-Pons}, Pilar and Siess, Lionel and Lattanzio, John C.},
  year = 2017,
  month = nov,
  journal = {PASA},
  volume = {34},
  pages = {e056},
  issn = {1323-3580},
  doi = {10.1017/pasa.2017.52},
  urldate = {2025-09-03}
}

@article{el-badry2023a,
  title = {The Fastest Stars in the {{Galaxy}}},
  author = {{El-Badry}, Kareem and Shen, Ken J. and Chandra, Vedant and Bauer, Evan B. and Fuller, Jim and Strader, Jay and Chomiuk, Laura and Naidu, Rohan P. and Caiazzo, Ilaria and Rodriguez, Antonio C. and Nagarajan, Pranav and Yamaguchi, Natsuko and Vanderbosch, Zachary P. and Roulston, Benjamin R. and G{\"a}nsicke, Boris and Han, Jiwon Jesse and Burdge, Kevin B. and Filippenko, Alexei V. and Brink, Thomas G. and Zheng, WeiKang},
  year = 2023,
  month = jul,
  journal = {Open J. Astrophys.},
  volume = {6},
  pages = {28},
  issn = {2565-6120},
  doi = {10.21105/astro.2306.03914},
  urldate = {2025-09-24}
}

@article{fink2010a,
  title = {Double-Detonation Sub-{{Chandrasekhar}} Supernovae: Can Minimum Helium Shell Masses Detonate the Core?},
  shorttitle = {Double-Detonation Sub-{{Chandrasekhar}} Supernovae},
  author = {Fink, M. and R{\"o}pke, F. K. and Hillebrandt, W. and Seitenzahl, I. R. and Sim, S. A. and Kromer, M.},
  year = 2010,
  month = may,
  journal = {A\&A},
  volume = {514},
  pages = {A53},
  issn = {0004-6361},
  doi = {10.1051/0004-6361/200913892},
  urldate = {2024-03-11}
}

@article{fink2014a,
  title = {Three-Dimensional Pure Deflagration Models with Nucleosynthesis and Synthetic Observables for {{Type Ia}} Supernovae},
  author = {Fink, Michael and Kromer, Markus and Seitenzahl, Ivo R. and {Ciaraldi-Schoolmann}, Franco and R{\"o}pke, Friedrich K. and Sim, Stuart A. and Pakmor, R{\"u}diger and Ruiter, Ashley J. and Hillebrandt, Wolfgang},
  year = 2014,
  month = feb,
  journal = {MNRAS},
  volume = {438},
  pages = {1762--1783},
  issn = {0035-8711},
  doi = {10.1093/mnras/stt2315},
  urldate = {2024-03-11}
}

@article{fink2018a,
  title = {Thermonuclear Explosions of Rapidly Differentially Rotating White Dwarfs: {{Candidates}} for Superluminous {{Type Ia}} Supernovae?},
  shorttitle = {Thermonuclear Explosions of Rapidly Differentially Rotating White Dwarfs},
  author = {Fink, M. and Kromer, M. and Hillebrandt, W. and R{\"o}pke, F. K. and Pakmor, R. and Seitenzahl, I. R. and Sim, S. A.},
  year = 2018,
  month = oct,
  journal = {A\&A},
  volume = {618},
  pages = {A124},
  issn = {0004-6361},
  doi = {10.1051/0004-6361/201833475},
  urldate = {2024-03-11}
}

@book{fryxell1989a,
  title = {Hydrodynamics and Nuclear Burning},
  author = {Fryxell, Bruce and M{\"u}ller, Ewald and Arnett, David},
  year = 1989,
  publisher = {Max-Planck-Inst. f{\"u}r Physik und Astrophysik}
}

@article{fuller1985a,
  title = {Stellar Weak Interaction Rates for Intermediate-Mass Nuclei. {{IV}} - {{Interpolation}} Procedures for Rapidly Varying Lepton Capture Rates Using Effective Log (Ft)-Values},
  author = {Fuller, G. M. and Fowler, W. A. and Newman, M. J.},
  year = 1985,
  month = jun,
  journal = {ApJ},
  volume = {293},
  pages = {1},
  issn = {0004-637X, 1538-4357},
  doi = {10.1086/163208},
  urldate = {2024-03-22},
  langid = {english}
}

@article{gutierrez1996a,
  title = {The {{Final Evolution}} of {{ONeMg Electron-Degenerate Cores}}},
  author = {Gutierrez, Jordi and {Garcia-Berro}, Enrique and Iben, Jr., Icko and Isern, Jordi and Labay, Javier and Canal, Ramon},
  year = 1996,
  month = mar,
  journal = {ApJ},
  volume = {459},
  pages = {701},
  issn = {0004-637X},
  doi = {10.1086/176934},
  urldate = {2025-10-07}
}

@article{iben1984a,
  title = {Supernovae of Type {{I}} as End Products of the Evolution of Binaries with Components of Moderate Initial Mass ({{M}} Not Greater than about 9 Solar Masses)},
  author = {Iben, Jr., I. and Tutukov, A. V.},
  year = 1984,
  month = feb,
  journal = {ApJS},
  volume = {54},
  pages = {335},
  issn = {0067-0049},
  doi = {10.1086/190932},
  urldate = {2025-02-05},
  langid = {english}
}

@article{isern1991a,
  title = {The {{Outcome}} of {{Explosive Ignition}} of {{ONeMg Cores}}: {{Supernovae}}, {{Neutron Stars}}, or ``{{Iron}}'' {{White Dwarfs}}?},
  shorttitle = {The {{Outcome}} of {{Explosive Ignition}} of {{ONeMg Cores}}},
  author = {Isern, Jordi and Canal, Ramon and Labay, Javier},
  year = 1991,
  month = may,
  journal = {ApJ},
  volume = {372},
  pages = {L83},
  issn = {0004-637X},
  doi = {10.1086/186029},
  urldate = {2025-10-07},
  langid = {english}
}

@article{itoh1996a,
  title = {Neutrino {{Energy Loss}} in {{Stellar Interiors}}. {{VII}}. {{Pair}}, {{Photo-}}, {{Plasma}}, {{Bremsstrahlung}}, and {{Recombination Neutrino Processes}}},
  author = {Itoh, Naoki and Hayashi, Hiroshi and Nishikawa, Akinori and Kohyama, Yasuharu},
  year = 1996,
  month = feb,
  journal = {ApJS},
  volume = {102},
  pages = {411},
  issn = {0067-0049},
  doi = {10.1086/192264},
  urldate = {2024-03-11}
}

@article{janka2008a,
  title = {Dynamics of Shock Propagation and Nucleosynthesis Conditions in {{O-Ne-Mg}} Core Supernovae},
  author = {Janka, H. -Th. and M{\"u}ller, B. and Kitaura, F. S. and Buras, R.},
  year = 2008,
  month = jul,
  journal = {A\&A},
  volume = {485},
  pages = {199--208},
  issn = {0004-6361},
  doi = {10.1051/0004-6361:20079334},
  urldate = {2025-09-08}
}

@misc{janka2025a,
  title = {Long-{{Term Multidimensional Models}} of {{Core-Collapse Supernovae}}: {{Progress}} and {{Challenges}}},
  shorttitle = {Long-{{Term Multidimensional Models}} of {{Core-Collapse Supernovae}}},
  author = {Janka, H.-Thomas},
  year = 2025,
  month = feb,
  number = {arXiv:2502.14836},
  eprint = {2502.14836},
  publisher = {arXiv},
  doi = {10.48550/arXiv.2502.14836},
  urldate = {2025-02-21},
  archiveprefix = {arXiv},
  langid = {english}
}

@article{jones2013a,
  title = {Advanced {{Burning Stages}} and {{Fate}} of 8-10 {{M}} {$\odot$} {{Stars}}},
  author = {Jones, S. and Hirschi, R. and Nomoto, K. and Fischer, T. and Timmes, F. X. and Herwig, F. and Paxton, B. and Toki, H. and Suzuki, T. and {Mart{\'i}nez-Pinedo}, G. and Lam, Y. H. and Bertolli, M. G.},
  year = 2013,
  month = aug,
  journal = {ApJ},
  volume = {772},
  pages = {150},
  issn = {0004-637X},
  doi = {10.1088/0004-637X/772/2/150},
  urldate = {2024-03-12}
}

@article{jones2016a,
  title = {Do Electron-Capture Supernovae Make Neutron Stars? - {{First}} Multidimensional Hydrodynamic Simulations of the Oxygen Deflagration},
  shorttitle = {Do Electron-Capture Supernovae Make Neutron Stars?},
  author = {Jones, S. and R{\"o}pke, F. K. and Pakmor, R. and Seitenzahl, I. R. and Ohlmann, S. T. and Edelmann, P. V. F.},
  year = 2016,
  month = sep,
  journal = {A\&A},
  volume = {593},
  pages = {A72},
  publisher = {EDP Sciences},
  issn = {0004-6361, 1432-0746},
  doi = {10.1051/0004-6361/201628321},
  urldate = {2024-01-08},
  copyright = {{\copyright} ESO, 2016},
  langid = {english}
}

@article{jones2019a,
  title = {Remnants and Ejecta of Thermonuclear Electron-Capture Supernovae: {{Constraining}} Oxygen-Neon Deflagrations in High-Density White Dwarfs},
  shorttitle = {Remnants and Ejecta of Thermonuclear Electron-Capture Supernovae},
  author = {Jones, S. and R{\"o}pke, F. K. and Fryer, C. and Ruiter, A. J. and Seitenzahl, I. R. and Nittler, L. R. and Ohlmann, S. T. and Reifarth, R. and Pignatari, M. and Belczynski, K.},
  year = 2019,
  month = feb,
  journal = {A\&A},
  volume = {622},
  pages = {A74},
  issn = {0004-6361, 1432-0746},
  doi = {10.1051/0004-6361/201834381},
  urldate = {2023-12-11}
}

@article{jones2019b,
  title = {A {{New Model}} for {{Electron-capture Supernovae}} in {{Galactic Chemical Evolution}}},
  author = {Jones, Samuel and C{\^o}t{\'e}, Benoit and R{\"o}pke, Friedrich K. and Wanajo, Shinya},
  year = 2019,
  month = sep,
  journal = {ApJ},
  volume = {882},
  number = {2},
  pages = {170},
  issn = {0004-637X, 1538-4357},
  doi = {10.3847/1538-4357/ab384e},
  urldate = {2024-05-31},
  langid = {english}
}

@article{kirsebom2019a,
  title = {Discovery of an {{Exceptionally Strong}} {$\beta$} -{{Decay Transition}} of {{F}} 20 and {{Implications}} for the {{Fate}} of {{Intermediate-Mass Stars}}},
  author = {Kirsebom, O. S. and Jones, S. and Str{\"o}mberg, D. F. and {Mart{\'i}nez-Pinedo}, G. and Langanke, K. and R{\"o}pke, F. K. and Brown, B. A. and Eronen, T. and Fynbo, H. O. U. and Hukkanen, M. and Idini, A. and Jokinen, A. and Kankainen, A. and Kostensalo, J. and Moore, I. and M{\"o}ller, H. and Ohlmann, S. T. and Penttil{\"a}, H. and Riisager, K. and {Rinta-Antila}, S. and Srivastava, P. C. and Suhonen, J. and Trzaska, W. H. and {\"A}yst{\"o}, J.},
  year = 2019,
  month = dec,
  journal = {Phys. Rev. Lett.},
  volume = {123},
  number = {26},
  pages = {262701},
  issn = {0031-9007, 1079-7114},
  doi = {10.1103/PhysRevLett.123.262701},
  urldate = {2025-04-25},
  langid = {english}
}

@article{kozyreva2021a,
  title = {Synthetic Observables for Electron-Capture Supernovae and Low-Mass Core Collapse Supernovae},
  author = {Kozyreva, Alexandra and Baklanov, Petr and Jones, Samuel and Stockinger, Georg and Janka, Hans-Thomas},
  year = 2021,
  month = may,
  journal = {MNRAS},
  volume = {503},
  pages = {797--814},
  publisher = {OUP},
  issn = {0035-8711},
  doi = {10.1093/mnras/stab350},
  urldate = {2024-07-17}
}

@article{kuhlen2006a,
  title = {Carbon {{Ignition}} in {{Type Ia Supernovae}}. {{II}}. {{A Three-dimensional Numerical Model}}},
  author = {Kuhlen, M. and Woosley, S. E. and Glatzmaier, G. A.},
  year = 2006,
  journal = {ApJ},
  volume = {640},
  pages = {407--416},
  issn = {0004-637X},
  doi = {10.1086/500105},
  urldate = {2025-07-30}
}

@article{lach2022a,
  title = {Models of Pulsationally Assisted Gravitationally Confined Detonations with Different Ignition Conditions},
  author = {Lach, F. and Callan, F. P. and Sim, S. A. and R{\"o}pke, F. K.},
  year = 2022,
  month = mar,
  journal = {A\&A},
  volume = {659},
  pages = {A27},
  publisher = {EDP Sciences},
  issn = {0004-6361, 1432-0746},
  doi = {10.1051/0004-6361/202142194},
  urldate = {2024-03-08},
  copyright = {{\copyright} ESO 2022},
  langid = {english}
}

@article{lach2022b,
  title = {Type {{Iax}} Supernovae from Deflagrations in {{Chandrasekhar}} Mass White Dwarfs},
  author = {Lach, F. and Callan, F. P. and Bubeck, D. and R{\"o}pke, F. K. and Sim, S. A. and Schrauth, M. and Ohlmann, S. T. and Kromer, M.},
  year = 2022,
  month = feb,
  journal = {A\&A},
  volume = {658},
  pages = {A179},
  publisher = {EDP Sciences},
  issn = {0004-6361, 1432-0746},
  doi = {10.1051/0004-6361/202141453},
  urldate = {2024-03-08},
  copyright = {{\copyright} ESO 2021},
  langid = {english}
}

@article{langanke2000a,
  title = {Shell-Model Calculations of Stellar Weak Interaction Rates: {{II}}. {{Weak}} Rates for Nuclei in the Mass Range /{{A}}=45-65 in Supernovae Environments},
  shorttitle = {Shell-Model Calculations of Stellar Weak Interaction Rates},
  author = {Langanke, K. and {Mart{\'i}nez-Pinedo}, G.},
  year = 2000,
  month = jun,
  journal = {Nucl. Phys. A},
  volume = {673},
  pages = {481--508},
  issn = {0375-9474},
  doi = {10.1016/S0375-9474(00)00131-7},
  urldate = {2023-09-14}
}

@article{liu2023a,
  title = {Type {{Ia Supernova Explosions}} in {{Binary Systems}}: {{A Review}}},
  shorttitle = {Type {{Ia Supernova Explosions}} in {{Binary Systems}}},
  author = {Liu, Zheng-Wei and R{\"o}pke, Friedrich K. and Han, Zhanwen},
  year = 2023,
  month = jul,
  journal = {Res. Astron. Astrophys.},
  volume = {23},
  number = {8},
  pages = {082001},
  issn = {1674-4527},
  doi = {10.1088/1674-4527/acd89e},
  urldate = {2025-02-05},
  langid = {english}
}

@article{marquardt2015a,
  title = {Type {{Ia}} Supernovae from Exploding Oxygen-Neon White Dwarfs},
  author = {Marquardt, Kai S. and Sim, Stuart A. and Ruiter, Ashley J. and Seitenzahl, Ivo R. and Ohlmann, Sebastian T. and Kromer, Markus and Pakmor, R{\"u}diger and R{\"o}pke, Friedrich K.},
  year = 2015,
  month = aug,
  journal = {A\&A},
  volume = {580},
  pages = {A118},
  issn = {0004-6361},
  doi = {10.1051/0004-6361/201525761},
  urldate = {2024-03-11}
}

@article{martinez-pinedo2014a,
  title = {Astrophysical Weak-Interaction Rates for Selected \${{A}}=20\$ and \${{A}}=24\$ Nuclei},
  author = {{Mart{\'i}nez-Pinedo}, G. and Lam, Y. H. and Langanke, K. and Zegers, R. G. T. and Sullivan, C.},
  year = 2014,
  month = apr,
  journal = {Phys. Rev. C},
  volume = {89},
  number = {4},
  pages = {045806},
  doi = {10.1103/PhysRevC.89.045806},
  urldate = {2025-10-07}
}

@article{miyaji1987a,
  title = {On the {{Collapse}} of 8--10 {{Msun Stars Due}} to {{Electron Capture}}},
  author = {Miyaji, Shigeki and Nomoto, Ken'ichi},
  year = 1987,
  month = jul,
  journal = {ApJ},
  volume = {318},
  pages = {307},
  issn = {0004-637X},
  doi = {10.1086/165368},
  urldate = {2025-09-03}
}

@article{nabi2004a,
  title = {Microscopic Calculations of Stellar Weak Interaction Rates and Energy Losses for Fp- and Fpg-Shell Nuclei},
  author = {Nabi, Jameel-Un and {Klapdor-Kleingrothaus}, Hans Volker},
  year = 2004,
  month = nov,
  journal = {At. Data Nucl. Data Tables},
  volume = {88},
  pages = {237--476},
  issn = {0092-640X},
  doi = {10.1016/j.adt.2004.09.002},
  urldate = {2023-09-14}
}

@article{nomoto1984a,
  title = {Evolution of 8-10 Solar Mass Stars toward Electron Capture Supernovae. {{I}} - {{Formation}} of Electron-Degenerate {{O}} + {{NE}} + {{MG}} Cores.},
  author = {Nomoto, K.},
  year = 1984,
  month = feb,
  journal = {ApJ},
  volume = {277},
  pages = {791--805},
  issn = {0004-637X},
  doi = {10.1086/161749},
  urldate = {2025-04-15}
}

@article{nomoto1987a,
  title = {Evolution of 8--10 {{Msun Stars}} toward {{Electron Capture Supernovae}}. {{II}}. {{Collapse}} of an {{O}} + {{NE}} + {{MG Core}}},
  author = {Nomoto, Ken'ichi},
  year = 1987,
  month = nov,
  journal = {ApJ},
  volume = {322},
  pages = {206},
  issn = {0004-637X},
  doi = {10.1086/165716},
  urldate = {2025-04-15}
}

@article{nomoto1991a,
  title = {Conditions for {{Accretion-induced Collapse}} of {{White Dwarfs}}},
  author = {Nomoto, Ken'ichi and Kondo, Yoji},
  year = 1991,
  month = jan,
  journal = {ApJ},
  volume = {367},
  pages = {L19},
  publisher = {IOP},
  issn = {0004-637X},
  doi = {10.1086/185922},
  urldate = {2025-04-15}
}

@article{nonaka2012a,
  title = {High-Resolution {{Simulations}} of {{Convection Preceding Ignition}} in {{Type Ia Supernovae Using Adaptive Mesh Refinement}}},
  author = {Nonaka, A. and Aspden, A. J. and Zingale, M. and Almgren, A. S. and Bell, J. B. and Woosley, S. E.},
  year = 2012,
  journal = {ApJ},
  volume = {745},
  pages = {73},
  issn = {0004-637X},
  doi = {10.1088/0004-637X/745/1/73},
  urldate = {2025-07-30}
}

@article{oda1994a,
  title = {Rate {{Tables}} for the {{Weak Processes}} of {\emph{Sd}}-{{Shell Nuclei}} in {{Stellar Matter}}},
  author = {Oda, T. and Hino, M. and Muto, K. and Takahara, M. and Sato, K.},
  year = 1994,
  month = mar,
  journal = {At. Data Nucl. Data Tables},
  volume = {56},
  number = {2},
  pages = {231--403},
  issn = {0092-640X},
  doi = {10.1006/adnd.1994.1007},
  urldate = {2024-03-22}
}

@article{ohlmann2014a,
  title = {The White Dwarf's Carbon Fraction as a Secondary Parameter of {{Type Ia}} Supernovae},
  author = {Ohlmann, Sebastian T. and Kromer, Markus and Fink, Michael and Pakmor, R{\"u}diger and Seitenzahl, Ivo R. and Sim, Stuart A. and R{\"o}pke, Friedrich K.},
  year = 2014,
  month = dec,
  journal = {A\&A},
  volume = {572},
  pages = {A57},
  issn = {0004-6361},
  doi = {10.1051/0004-6361/201423924},
  urldate = {2024-03-11}
}

@article{osher1988a,
  title = {Fronts {{Propagating}} with {{Curvature-Dependent Speed}}: {{Algorithms Based}} on {{Hamilton-Jacobi Formulations}}},
  shorttitle = {Fronts {{Propagating}} with {{Curvature-Dependent Speed}}},
  author = {Osher, Stanley and Sethian, James A.},
  year = 1988,
  month = nov,
  journal = {J. Comp. Phys.},
  volume = {79},
  pages = {12--49},
  issn = {0021-9991},
  doi = {10.1016/0021-9991(88)90002-2},
  urldate = {2024-03-11}
}

@article{pakmor2012a,
  title = {Stellar {{GADGET}}: A Smoothed Particle Hydrodynamics Code for Stellar Astrophysics and Its Application to {{Type Ia}} Supernovae from White Dwarf Mergers},
  shorttitle = {Stellar {{GADGET}}},
  author = {Pakmor, R. and Edelmann, P. and R{\"o}pke, F. K. and Hillebrandt, W.},
  year = 2012,
  month = aug,
  journal = {MNRAS},
  volume = {424},
  pages = {2222--2231},
  issn = {0035-8711},
  doi = {10.1111/j.1365-2966.2012.21383.x},
  urldate = {2023-09-14}
}

@article{paxton2011a,
  title = {Modules for {{Experiments}} in {{Stellar Astrophysics}} ({{MESA}})},
  author = {Paxton, Bill and Bildsten, Lars and Dotter, Aaron and Herwig, Falk and Lesaffre, Pierre and Timmes, Frank},
  year = 2011,
  month = jan,
  journal = {ApJS},
  volume = {192},
  pages = {3},
  issn = {0067-0049},
  doi = {10.1088/0067-0049/192/1/3},
  urldate = {2024-03-12}
}

@article{paxton2013a,
  title = {Modules for {{Experiments}} in {{Stellar Astrophysics}} ({{MESA}}): {{Planets}}, {{Oscillations}}, {{Rotation}}, and {{Massive Stars}}},
  shorttitle = {Modules for {{Experiments}} in {{Stellar Astrophysics}} ({{MESA}})},
  author = {Paxton, Bill and Cantiello, Matteo and Arras, Phil and Bildsten, Lars and Brown, Edward F. and Dotter, Aaron and Mankovich, Christopher and Montgomery, M. H. and Stello, Dennis and Timmes, F. X. and Townsend, Richard},
  year = 2013,
  month = sep,
  journal = {ApJS},
  volume = {208},
  pages = {4},
  issn = {0067-0049},
  doi = {10.1088/0067-0049/208/1/4},
  urldate = {2024-03-12}
}

@article{paxton2015a,
  title = {Modules for {{Experiments}} in {{Stellar Astrophysics}} ({{MESA}}): {{Binaries}}, {{Pulsations}}, and {{Explosions}}},
  shorttitle = {Modules for {{Experiments}} in {{Stellar Astrophysics}} ({{MESA}})},
  author = {Paxton, Bill and Marchant, Pablo and Schwab, Josiah and Bauer, Evan B. and Bildsten, Lars and Cantiello, Matteo and Dessart, Luc and Farmer, R. and Hu, H. and Langer, N. and Townsend, R. H. D. and Townsley, Dean M. and Timmes, F. X.},
  year = 2015,
  month = sep,
  journal = {ApJS},
  volume = {220},
  pages = {15},
  issn = {0067-0049},
  doi = {10.1088/0067-0049/220/1/15},
  urldate = {2024-03-12}
}

@article{podsiadlowski2004a,
  title = {The {{Effects}} of {{Binary Evolution}} on the {{Dynamics}} of {{Core Collapse}} and {{Neutron Star Kicks}}},
  author = {Podsiadlowski, {\relax Ph}. and Langer, N. and Poelarends, A. J. T. and Rappaport, S. and Heger, A. and Pfahl, E.},
  year = 2004,
  month = sep,
  journal = {ApJ},
  volume = {612},
  pages = {1044--1051},
  issn = {0004-637X},
  doi = {10.1086/421713},
  urldate = {2025-09-04}
}

@article{potekhin2000a,
  title = {Equation of State of Fully Ionized Electron-Ion Plasmas. {{II}}. {{Extension}} to Relativistic Densities and to the Solid Phase},
  author = {Potekhin, Alexander Y. and Chabrier, Gilles},
  year = 2000,
  month = dec,
  journal = {Phys. Rev. E},
  volume = {62},
  pages = {8554--8563},
  issn = {1063-651X},
  doi = {10.1103/PhysRevE.62.8554},
  urldate = {2024-03-11}
}

@article{raddi2019a,
  title = {Partly Burnt Runaway Stellar Remnants from Peculiar Thermonuclear Supernovae},
  author = {Raddi, R and Hollands, M A and Koester, D and Hermes, J J and G{\"a}nsicke, B T and Heber, U and Shen, K J and Townsley, D M and Pala, A F and Reding, J S and Toloza, O F and Pelisoli, I and Geier, S and Gentile~Fusillo, N P and Munari, U and Strader, J},
  year = 2019,
  month = oct,
  journal = {MNRAS},
  volume = {489},
  number = {2},
  pages = {1489--1508},
  issn = {0035-8711},
  doi = {10.1093/mnras/stz1618},
  urldate = {2025-10-17}
}

@article{reinecke1999a,
  title = {Thermonuclear Explosions of {{Chandrasekhar-mass C}}+{{O}} White Dwarfs},
  author = {Reinecke, M. and Hillebrandt, W. and Niemeyer, J. C.},
  year = 1999,
  month = jul,
  journal = {A\&A},
  volume = {347},
  pages = {739--747},
  issn = {0004-6361},
  doi = {10.48550/arXiv.astro-ph/9812120},
  urldate = {2024-03-11}
}

@article{reinecke1999b,
  title = {A New Model for Deflagration Fronts in Reactive Fluids},
  author = {Reinecke, M. and Hillebrandt, W. and Niemeyer, J. C. and Klein, R. and Gr{\"o}bl, A.},
  year = 1999,
  month = jul,
  journal = {A\&A},
  volume = {347},
  pages = {724--733},
  issn = {0004-6361},
  doi = {10.48550/arXiv.astro-ph/9812119},
  urldate = {2024-03-11}
}

@article{ritossa1999a,
  title = {On the {{Evolution}} of {{Stars}} That {{Form Electron-degenerate Cores Processed}} by {{Carbon Burning}}. {{V}}. {{Shell Convection Sustained}} by {{Helium Burning}}, {{Transient Neon Burning}}, {{Dredge-out}}, {{Urca Cooling}}, and {{Other Properties}} of an 11 {{Msolar Population I Model Star}}},
  author = {Ritossa, Claudio and {Garc{\'i}a-Berro}, Enrique and Iben, Jr., Icko},
  year = 1999,
  month = apr,
  journal = {ApJ},
  volume = {515},
  pages = {381--397},
  issn = {0004-637X},
  doi = {10.1086/307017},
  urldate = {2025-09-04}
}

@article{ropke2005a,
  title = {Full-Star Type {{Ia}} Supernova Explosion Models},
  author = {R{\"o}pke, F. K. and Hillebrandt, W.},
  year = 2005,
  month = feb,
  journal = {A\&A},
  volume = {431},
  pages = {635--645},
  issn = {0004-6361},
  doi = {10.1051/0004-6361:20041859},
  urldate = {2024-03-11}
}

@article{ropke2006a,
  title = {Multi-Spot Ignition in Type {{Ia}} Supernova Models},
  author = {R{\"o}pke, F. K. and Hillebrandt, W. and Niemeyer, J. C. and Woosley, S. E.},
  year = 2006,
  month = mar,
  journal = {A\&A},
  volume = {448},
  pages = {1--14},
  issn = {0004-6361},
  doi = {10.1051/0004-6361:20053926},
  urldate = {2024-03-11}
}

@article{ropke2007a,
  title = {A {{Three}}-{{Dimensional Deflagration Model}} for {{Type Ia Supernovae Compared}} with {{Observations}}},
  author = {R{\"o}pke, F. K. and Hillebrandt, W. and Schmidt, W. and Niemeyer, J. C. and Blinnikov, S. I. and Mazzali, P. A.},
  year = 2007,
  month = oct,
  journal = {ApJ},
  volume = {668},
  number = {2},
  pages = {1132--1139},
  issn = {0004-637X, 1538-4357},
  doi = {10.1086/521347},
  urldate = {2024-03-11},
  langid = {english}
}

@article{rusanov1962a,
  title = {The Calculation of the Interaction of Non-Stationary Shock Waves and Obstacles},
  author = {Rusanov, V. V},
  year = 1962,
  month = jan,
  journal = {USSR Comput. Math. \& Math. Phys.},
  volume = {1},
  number = {2},
  pages = {304--320},
  issn = {0041-5553},
  doi = {10.1016/0041-5553(62)90062-9},
  urldate = {2025-05-01}
}

@article{schmidt2006a,
  title = {A Localised Subgrid Scale Model for Fluid Dynamical Simulations in Astrophysics - {{I}}. {{Theory}} and Numerical Tests},
  author = {Schmidt, W. and Niemeyer, J. C. and Hillebrandt, W.},
  year = 2006,
  month = apr,
  journal = {A\&A},
  volume = {450},
  number = {1},
  pages = {265--281},
  issn = {0004-6361, 1432-0746},
  doi = {10.1051/0004-6361:20053617},
  urldate = {2024-11-22},
  copyright = {{\copyright} ESO, 2006},
  langid = {english}
}

@article{schmidt2006b,
  title = {A Localised Subgrid Scale Model for Fluid Dynamical Simulations in Astrophysics. {{II}}. {{Application}} to Type {{Ia}} Supernovae},
  author = {Schmidt, W. and Niemeyer, J. C. and Hillebrandt, W. and R{\"o}pke, F. K.},
  year = 2006,
  month = apr,
  journal = {A\&A},
  volume = {450},
  pages = {283--294},
  issn = {0004-6361},
  doi = {10.1051/0004-6361:20053618},
  urldate = {2024-03-11}
}

@article{schwab2015a,
  title = {Thermal Runaway during the Evolution of {{ONeMg}} Cores towards Accretion-Induced Collapse},
  author = {Schwab, Josiah and Quataert, Eliot and Bildsten, Lars},
  year = 2015,
  month = oct,
  journal = {MNRAS},
  volume = {453},
  pages = {1910--1927},
  issn = {0035-8711},
  doi = {10.1093/mnras/stv1804},
  urldate = {2025-09-04}
}

@article{schwab2017a,
  title = {The Importance of {{Urca-process}} Cooling in Accreting {{ONe}} White Dwarfs},
  author = {Schwab, Josiah and Bildsten, Lars and Quataert, Eliot},
  year = 2017,
  month = dec,
  journal = {MNRAS},
  volume = {472},
  number = {3},
  pages = {3390--3406},
  issn = {0035-8711},
  doi = {10.1093/mnras/stx2169},
  urldate = {2025-10-07}
}

@article{schwab2019b,
  title = {Residual {{Carbon}} in {{Oxygen}}--{{Neon White Dwarfs}} and {{Its Implications}} for {{Accretion-induced Collapse}}},
  author = {Schwab, Josiah and Rocha, Kyle Akira},
  year = 2019,
  month = feb,
  journal = {ApJ},
  volume = {872},
  number = {2},
  pages = {131},
  issn = {0004-637X, 1538-4357},
  doi = {10.3847/1538-4357/aaffdc},
  urldate = {2025-08-19},
  langid = {english}
}

@article{schwab2020a,
  title = {Laminar {{Flame Speeds}} in {{Degenerate Oxygen}}--{{Neon Mixtures}}},
  author = {Schwab, Josiah and Farmer, R. and Timmes, F. X.},
  year = 2020,
  month = feb,
  journal = {ApJ},
  volume = {891},
  number = {1},
  pages = {5},
  publisher = {The American Astronomical Society},
  issn = {0004-637X},
  doi = {10.3847/1538-4357/ab6f03},
  urldate = {2024-03-08},
  langid = {english}
}

@article{seitenzahl2009a,
  title = {Nuclear Statistical Equilibrium for {{Type Ia}} Supernova Simulations},
  author = {Seitenzahl, Ivo R. and Townsley, Dean M. and Peng, Fang and Truran, James W.},
  year = 2009,
  month = jan,
  journal = {At. Data Nucl. Data Tables},
  volume = {95},
  number = {1},
  pages = {96--114},
  issn = {0092640X},
  doi = {10.1016/j.adt.2008.08.001},
  urldate = {2024-01-12},
  langid = {english}
}

@article{seitenzahl2013a,
  title = {Three-Dimensional Delayed-Detonation Models with Nucleosynthesis for {{Type Ia}} Supernovae},
  author = {Seitenzahl, Ivo R. and {Ciaraldi-Schoolmann}, Franco and R{\"o}pke, Friedrich K. and Fink, Michael and Hillebrandt, Wolfgang and Kromer, Markus and Pakmor, R{\"u}diger and Ruiter, Ashley J. and Sim, Stuart A. and Taubenberger, Stefan},
  year = 2013,
  month = feb,
  journal = {MNRAS},
  volume = {429},
  pages = {1156--1172},
  issn = {0035-8711},
  doi = {10.1093/mnras/sts402},
  urldate = {2024-03-11}
}

@article{seitenzahl2015a,
  title = {Neutrino and Gravitational Wave Signal of a Delayed-Detonation Model of Type {{Ia}} Supernovae},
  author = {Seitenzahl, Ivo R. and Herzog, Matthias and Ruiter, Ashley J. and Marquardt, Kai and Ohlmann, Sebastian T. and R{\"o}pke, Friedrich K.},
  year = 2015,
  month = dec,
  journal = {Phys. Rev. D},
  volume = {92},
  pages = {124013},
  issn = {1550-79980556-2821},
  doi = {10.1103/PhysRevD.92.124013},
  urldate = {2024-03-11}
}

@article{sullivan2015a,
  title = {{{THE SENSITIVITY OF CORE-COLLAPSE SUPERNOVAE TO NUCLEAR ELECTRON CAPTURE}}},
  author = {Sullivan, Chris and O'Connor, Evan and Zegers, Remco G. T. and Grubb, Thomas and Austin, Sam M.},
  year = 2015,
  month = dec,
  journal = {ApJ},
  volume = {816},
  number = {1},
  pages = {44},
  publisher = {The American Astronomical Society},
  issn = {0004-637X},
  doi = {10.3847/0004-637X/816/1/44},
  urldate = {2023-12-05},
  langid = {english}
}

@article{suzuki2024a,
  title = {Neutrinos from {{Core-Collapse Supernova Explosions}}},
  author = {Suzuki, Hideyuki},
  year = 2024,
  month = may,
  journal = {Prog. Theor. Exp. Phys.},
  volume = {2024},
  number = {5},
  pages = {05B101},
  issn = {2050-3911},
  doi = {10.1093/ptep/ptae056},
  urldate = {2025-10-17}
}

@article{tauris2015a,
  title = {Ultra-Stripped Supernovae: Progenitors and Fate},
  shorttitle = {Ultra-Stripped Supernovae},
  author = {Tauris, Thomas M. and Langer, Norbert and Podsiadlowski, Philipp},
  year = 2015,
  month = aug,
  journal = {MNRAS},
  volume = {451},
  pages = {2123--2144},
  issn = {0035-8711},
  doi = {10.1093/mnras/stv990},
  urldate = {2025-09-04}
}

@article{timmes1992a,
  title = {The {{Conductive Propagation}} of {{Nuclear Flames}}. {{I}}. {{Degenerate C}} + {{O}} and {{O}} + {{NE}} + {{MG White Dwarfs}}},
  author = {Timmes, F. X. and Woosley, S. E.},
  year = 1992,
  month = sep,
  journal = {ApJ},
  volume = {396},
  pages = {649},
  issn = {0004-637X},
  doi = {10.1086/171746},
  urldate = {2024-03-11}
}

@article{timmes1999a,
  title = {The {{Accuracy}}, {{Consistency}}, and {{Speed}} of {{Five Equations}} of {{State}} for {{Stellar Hydrodynamics}}},
  author = {Timmes, F. X. and Arnett, Dave},
  year = 1999,
  month = nov,
  journal = {ApJS},
  volume = {125},
  pages = {277--294},
  issn = {0067-0049},
  doi = {10.1086/313271},
  urldate = {2024-03-11}
}

@article{timmes2000a,
  title = {The {{Accuracy}}, {{Consistency}}, and {{Speed}} of an {{Electron-Positron Equation}} of {{State Based}} on {{Table Interpolation}} of the {{Helmholtz Free Energy}}},
  author = {Timmes, F. X. and Swesty, F. Douglas},
  year = 2000,
  month = feb,
  journal = {ApJS},
  volume = {126},
  pages = {501--516},
  issn = {0067-0049},
  doi = {10.1086/313304},
  urldate = {2023-09-13}
}

@article{vennes2017a,
  title = {An Unusual White Dwarf Star May Be a Surviving Remnant of a Subluminous {{Type Ia}} Supernova},
  author = {Vennes, S. and Nemeth, P. and Kawka, A. and Thorstensen, J. R. and Khalack, V. and Ferrario, L. and Alper, E. H.},
  year = 2017,
  month = aug,
  journal = {Science},
  doi = {10.1126/science.aam8378},
  urldate = {2025-10-17},
  copyright = {Copyright {\copyright} 2017 The Authors, some rights reserved; exclusive licensee American Association for the Advancement of Science. No claim to original U.S. Government Works},
  langid = {english}
}

@article{wanajo2011a,
  title = {Electron-Capture {{Supernovae}} as {{The Origin}} of {{Elements Beyond Iron}}},
  author = {Wanajo, Shinya and Janka, Hans-Thomas and M{\"u}ller, Bernhard},
  year = 2011,
  month = jan,
  journal = {ApJ},
  volume = {726},
  pages = {L15},
  issn = {0004-637X},
  doi = {10.1088/2041-8205/726/2/L15},
  urldate = {2025-09-08}
}

@article{whelan1973a,
  title = {Binaries and {{Supernovae}} of {{Type I}}},
  author = {Whelan, John and Iben, Jr., Icko},
  year = 1973,
  month = dec,
  journal = {ApJ},
  volume = {186},
  pages = {1007--1014},
  issn = {0004-637X},
  doi = {10.1086/152565},
  urldate = {2025-02-05}
}

@article{yakovlev1989a,
  title = {Degenerate {{Cores}} of {{White Dwarfs}} and {{Envelopes}} of {{Neutron Stars}} - {{Thermodynamics}} and {{Plasma Screening}} in {{Thermonuclear Reactions}}},
  author = {Yakovlev, D. G. and Shalybkov, D. A.},
  year = 1989,
  month = jan,
  journal = {Astrophys.{\textasciitilde}Space{\textasciitilde}Phys.{\textasciitilde}Rev.},
  volume = {7},
  pages = {311},
  urldate = {2024-03-11}
}

@article{zingale2009a,
  title = {Low {{Mach Number Modeling}} of {{Type IA Supernovae}}. {{IV}}. {{White Dwarf Convection}}},
  author = {Zingale, M. and Almgren, A. S. and Bell, J. B. and Nonaka, A. and Woosley, S. E.},
  year = 2009,
  journal = {ApJ},
  volume = {704},
  pages = {196--210},
  issn = {0004-637X},
  doi = {10.1088/0004-637X/704/1/196},
  urldate = {2025-07-30}
}

@article{takahashi2019a,
  title = {The {{Evolution}} toward {{Electron Capture Supernovae}}: {{The Flame Propagation}} and the {{Pre-bounce Electron}}--{{Neutrino Radiation}}},
  shorttitle = {The {{Evolution}} toward {{Electron Capture Supernovae}}},
  author = {Takahashi, Koh and Sumiyoshi, Kohsuke and Yamada, Shoichi and Umeda, Hideyuki and Yoshida, Takashi},
  year = 2019,
  month = jan,
  journal = {ApJ},
  volume = {871},
  number = {2},
  pages = {153},
  issn = {0004-637X},
  doi = {10.3847/1538-4357/aaf8a8},
  urldate = {2025-12-08},
  langid = {english}
}

@article{zha2022a,
  title = {Hydrodynamic Simulations of Electron-Capture Supernovae: Progenitor and Dimension Dependence},
  shorttitle = {Hydrodynamic Simulations of Electron-Capture Supernovae},
  author = {Zha, Shuai and O'Connor, Evan P and Couch, Sean M and Leung, Shing-Chi and Nomoto, Ken'ichi},
  year = 2022,
  month = jun,
  journal = {MNRAS},
  volume = {513},
  number = {1},
  pages = {1317--1328},
  issn = {0035-8711},
  doi = {10.1093/mnras/stac1035},
  urldate = {2025-12-08}
}

@article{zha2019a,
  title = {Evolution of {{ONeMg Core}} in {{Super-AGB Stars}} toward {{Electron-capture Supernovae}}: {{Effects}} of {{Updated Electron-capture Rate}}},
  shorttitle = {Evolution of {{ONeMg Core}} in {{Super-AGB Stars}} toward {{Electron-capture Supernovae}}},
  author = {Zha, Shuai and Leung, Shing-Chi and Suzuki, Toshio and Nomoto, Ken'ichi},
  year = 2019,
  month = nov,
  journal = {ApJ},
  volume = {886},
  number = {1},
  pages = {22},
  issn = {0004-637X},
  doi = {10.3847/1538-4357/ab4b4b},
  urldate = {2025-12-08},
  langid = {english}
}

@article{leung2020a,
  title = {Electron-Capture {{Supernovae}} of {{Super-AGB Stars}}: {{Sensitivity}} on {{Input Physics}}},
  shorttitle = {Electron-Capture {{Supernovae}} of {{Super-AGB Stars}}},
  author = {Leung, Shing-Chi and Nomoto, Ken'ichi and Suzuki, Tomoharu},
  year = 2020,
  month = jan,
  journal = {ApJ},
  volume = {889},
  number = {1},
  pages = {34},
  issn = {0004-637X},
  doi = {10.3847/1538-4357/ab5d2f},
  urldate = {2025-12-08},
  langid = {english}
}

@article{pepper2022a,
  title = {The Impact of the Uncertainties in the {{12C}}({$\alpha$},{$\gamma$}){{16O}} Reaction Rate on the Evolution of Low- to Intermediate-Mass Stars},
  author = {Pepper, Ben T. and Istrate, A. G. and Romero, A. D. and Kepler, S. O.},
  year = 2022,
  month = apr,
  journal = {MNRAS},
  volume = {513},
  number = {1},
  pages = {1499--1512},
  issn = {0035-8711, 1365-2966},
  doi = {10.1093/mnras/stac1016},
  urldate = {2025-12-09}
}

@article{an2015a,
  title = {Astrophysical {{S}} Factor of the {{C}} 12 ( {$\alpha$} , {$\gamma$} ) {{O}} 16 Reaction Calculated with Reduced {{R}} -Matrix Theory},
  author = {An, Zhen-Dong and Chen, Zhen-Peng and Ma, Yu-Gang and Yu, Jian-Kai and Sun, Ye-Ying and Fan, Gong-Tao and Li, Yong-Jiang and Xu, Hang-Hua and Huang, Bo-Song and Wang, Kan},
  year = 2015,
  month = oct,
  journal = {Phys. Rev. C},
  volume = {92},
  number = {4},
  pages = {045802},
  issn = {0556-2813, 1089-490X},
  doi = {10.1103/PhysRevC.92.045802},
  urldate = {2025-12-09},
  copyright = {http://link.aps.org/licenses/aps-default-license},
  langid = {english}
}

\onecolumn
\begin{appendix}
\section{Additional figures and tables}
This section contains supplementary figures and tables containing various properties of all simulated models.
Figure~\ref{fig:protocol2} shows a continuation of Figure~\ref{fig:protocol1} for larger ignition radii.
Table~\ref{tab:model_list} lists all the simulated models, as well as their predicted outcome, that is, explosion or collapse.

\begin{figure*}[h!]
    \centering
    \includegraphics[width=\linewidth,keepaspectratio]{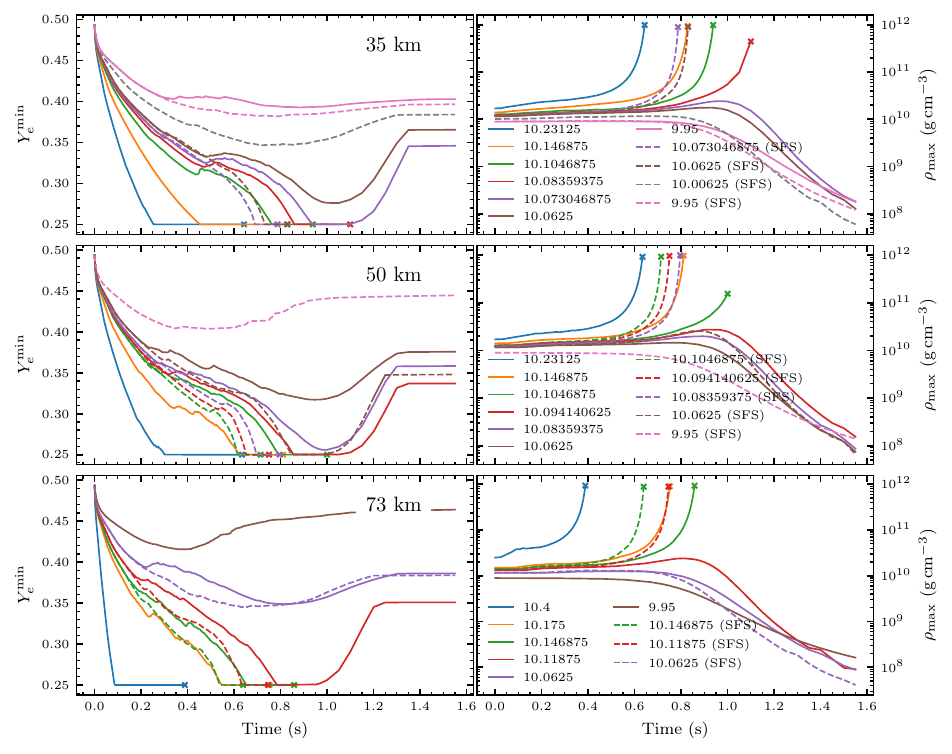}
    \caption{Continuation of Figure~\ref{fig:protocol1}.}
    \label{fig:protocol2}
\end{figure*}

\begin{longtable}{l|cccc}
\caption{List of simulated models and their initial properties as well as the simulation outcome for both choices of laminar flame speed parameterization.\label{tab:model_list}}\\
\hline\hline
Model & $\log\rhoini$ & $r_\mathrm{ign}$ (km) & \citetalias{timmes1992a} & \citetalias{schwab2020a}\\
\hline
\endfirsthead
\caption{Continued.}\\
\hline\hline
Model & $\log\rhoini$ & $r_\mathrm{ign}$ (km) & \citetalias{timmes1992a} & \citetalias{schwab2020a}\\
\hline
\endhead
\hline
\endfoot
rho9.95\_r2.5(\_sfs) & 9.95 & 2.5 & Expl. & Expl. \\
rho9.95\_r10(\_sfs) & 9.95 & 10 & Expl. & Expl. \\
rho9.95\_r20\_sfs & 9.95 & 20 & ... & Expl. \\
rho9.95\_r35(\_sfs) & 9.95 & 35 & Expl. & Expl. \\
rho9.95\_r50\_sfs & 9.95 & 50 & ... & Expl. \\
rho9.95\_r73 & 9.95 & 73 & Expl. & ... \\
rho10.00625\_r2.5(\_sfs) & 10.00625 & 2.5 & Expl. & Coll. \\
rho10.00625\_r10(\_sfs) & 10.00625 & 10 & Expl. & Coll. \\
rho10.00625\_r35\_sfs & 10.00625 & 35 & ... & Expl. \\
rho10.0203125\_r10(\_sfs) & 10.0203125 & 10 & Expl. & Coll. \\
rho10.0203125\_r20(\_sfs) & 10.0203125 & 20 & Expl. & Coll. \\
rho10.034375\_r2.5(\_sfs) & 10.034375 & 2.5 & Coll. & Coll. \\
rho10.034375\_r10(\_sfs) & 10.034375 & 10 & Coll. & Coll. \\
rho10.05195312\_r20 & 10.05195312 & 20 & Expl. & ... \\
rho10.0625\_r2.5 & 10.0625 & 2.5 & Coll. & ... \\
rho10.0625\_r10 & 10.0625 & 10 & Coll. & ... \\
rho10.0625\_r35(\_sfs) & 10.0625 & 35 & Expl. & Coll. \\
rho10.0625\_r50(\_sfs) & 10.0625 & 50 & Expl. & Expl. \\
rho10.0625\_r73(\_sfs) & 10.0625 & 73 & Expl. & Expl. \\
rho10.06777344\_r20 & 10.06777344 & 20 & Coll. & ... \\
rho10.073046875\_r35(\_sfs) & 10.073046875 & 35 & Expl. & Coll. \\
rho10.08359375\_r20 & 10.08359375 & 20 & Coll. & ... \\
rho10.08359375\_r35 & 10.08359375 & 35 & Coll. & ... \\
rho10.08359375\_r50(\_sfs) & 10.08359375 & 50 & Expl. & Coll. \\
rho10.094140625\_r50(\_sfs) & 10.094140625 & 50 & Expl. & Coll. \\
rho10.1046875\_r35 & 10.1046875 & 35 & Coll. & ... \\
rho10.1046875\_r50(\_sfs) & 10.1046875 & 50 & Coll. & Coll. \\
rho10.11875\_r73(\_sfs) & 10.11875 & 73 & Expl. & Coll. \\
rho10.146875\_r35 & 10.146875 & 35 & Coll. & ... \\
rho10.146875\_r50 & 10.146875 & 50 & Coll. & ... \\
rho10.146875\_r73(\_sfs) & 10.146875 & 73 & Coll. & Coll. \\
rho10.175\_r2.5 & 10.175 & 2.5 & Coll. & ... \\
rho10.175\_r73 & 10.175 & 73 & Coll. & ... \\
rho10.23125\_r35 & 10.23125 & 35 & Coll. & ... \\
rho10.23125\_r50 & 10.23125 & 50 & Coll. & ... \\
rho10.4\_r2.5 & 10.4 & 2.5 & Coll. & ... \\
rho10.4\_r10 & 10.4 & 10 & Coll. & ... \\
rho10.4\_r73 & 10.4 & 73 & Coll. & ... \\
\end{longtable}
\end{appendix}

\end{document}